\RequirePackage{fix-cm}

\documentclass[smallcondensed
]{svjour3}     
\smartqed  
\usepackage{graphicx}
\usepackage{epstopdf}
\usepackage{geometry}
\usepackage{comment}
\usepackage{xspace}
\usepackage{amsmath}
\usepackage{amssymb}

\usepackage{epsfig}
\usepackage{subfigure}
\usepackage{algorithmic}
\usepackage{graphics}
\usepackage{alltt}
\usepackage{float}
\usepackage{array}
\usepackage{footmisc}
\usepackage{comment}
\usepackage{algorithm}
\usepackage{url}
\usepackage[misc]{ifsym}

\usepackage[justification=centering]{caption}

\begin{document}

\title{SVS-JOIN: Efficient Spatial Visual Similarity Join for Mutilmedia Data
}


\author{Chengyuan Zhang  $^\dagger$\and
        Ruipeng Chen $^\dagger$\and
        Lei Zhu $^\dagger$\and
        Zuping Zhang $^\dagger$\and
        Fang Huang $^\dagger$\and
        Yunwu Lin $^\dagger$\and
        }


\institute{Chengyuan Zhang \at
            \email{cyzhang@csu.edu.cn}
            \and
            Ruipeng Chen \at
              \email{rpchen@csu.edu.cn}
           \and
           \Letter Lei Zhu \at
              \email{leizhu@csu.edu.cn}
           \and
           Zuping Zhang \at
              \email{zpzhang@csu.edu.cn}
           \and
           Fang Huang \at
              \email{hfang@csu.edu.cn}
           \and
           Yunwu Lin \at
              \email{lywcsu@csu.edu.cn}
           \and
           $^\dagger$ School of Information Science and Engineering, Central South University, PR China\\
}

\date{Received: date / Accepted: date}

\maketitle

\begin{abstract}
In the big data era, massive amount of multimedia data with geo-tags has been generated and collected by mobile smart devices equipped with mobile communications module and position sensor module. This trend has put forward higher request on large-scale of geo-multimedia data retrieval. Spatial similarity join is one of the important problem in the area of spatial database. Previous works focused on textual document with geo-tags, rather than geo-multimedia data such as geo-images. In this paper, we study a novel search problem named spatial visual similarity join (SVS-JOIN for short), which aims to find similar geo-image pairs in both the aspects of geo-location and visual content. We propose the definition of SVS-JOIN at the first time and present how to measure geographical similarity and visual similarity. Then we introduce a baseline inspired by the method for textual similarity join and a extension named SVS-JOIN$_G$ which applies spatial grid strategy to improve the efficiency. To further improve the performance of search, we develop a novel approach called SVS-JOIN$_Q$ which utilizes a quadtree and a global inverted index. Experimental evaluations on real geo-image datasets demonstrate that our solution has a really high performance.


\end{abstract}

\section{Introduction}
\label{intro}

In the big data era, Internet techniques and claud services such as online social networking services, search engine and multimedia sharing services are developing rapidly, which generate and storage large-scale of multimedia data~\cite{DBLP:conf/cikm/WangLZ13,InfYang13,KAISYang16,DBLP:conf/mm/WangLWZZ14,DBLP:conf/mm/WangLWZ15,DBLP:journals/tip/WangLWZ17}, e.g., text, image, audio and video. For example, More and more people use online social networking services such as Facebook (https://facebook.com/), Twitter (http://www.twitter.com/), Linkedin (https://www.linkedin.com/), Weibo (https://weibo.com/), etc. to making friends and sharing their hobbies or work information by uploading texts, images, or short videos. On the other hand, for multimedia data~\cite{DBLP:journals/corr/abs-1804-11013} sharing services, such as Flickr(https://www.flickr.com/), the most famous photo sharing web site, more than 3.5 million new images uploaded to it everyday in March 2013. In addition, every minute there are 100 hours of videos are uploaded to YouTube (https://www.youtube.com/), the largest video sharing service all around the world, and more than 2 billion videos totally stored in this platform by the end of 2013. In China, IQIYI (http://www.iqiyi.com/) is the largest video sharing web site. the total watch time monthly of this online video service exceeded 42 billion minutes. These multimedia web services not only provide great convenience for our daily life, but creates possibilities for the generation, collection, storage and sharing of large-scale multimedia data~\cite{DBLP:conf/sigir/WangLWZZ15,DBLP:journals/tip/WangLWZZH15,DBLP:conf/ijcai/WangZWLFP16}. Moreover, this trend has put forward greater challenges for massive multimedia data retrieval~\cite{DBLP:journals/cviu/WuWGHL18,DBLP:journals/corr/abs-1708-02288,DBLP:conf/pakdd/WangLZW14}.

Mobile smart devices equipped with mobile communications module (e.g., WiFi and 4G module) and position sensor module (e.g., GPS-Module) such as smartphones and tablets collective huge amounts of multimedia data~\cite{DBLP:journals/pr/WuWGL18,DBLP:journals/tnn/WangZWLZ17} with geo-tags. For example, users can take photos or videos~\cite{NNLS2018} with the geo-location information of the shoot place. Many mobile applications such as WeiChat, Twitter and Instagram support uploading and sharing of images and text with geo-tags. Other location-based services such as Google Places, Yahoo!Local, and Dianping provide the query services for geo-multimedia data by taking into account both geographical proximity and multimedia data similarity.

\begin{figure*}
\newskip\subfigtoppskip \subfigtopskip = -0.1cm
\begin{minipage}[b]{0.99\linewidth}
\begin{center}
     \includegraphics[width=1\linewidth]{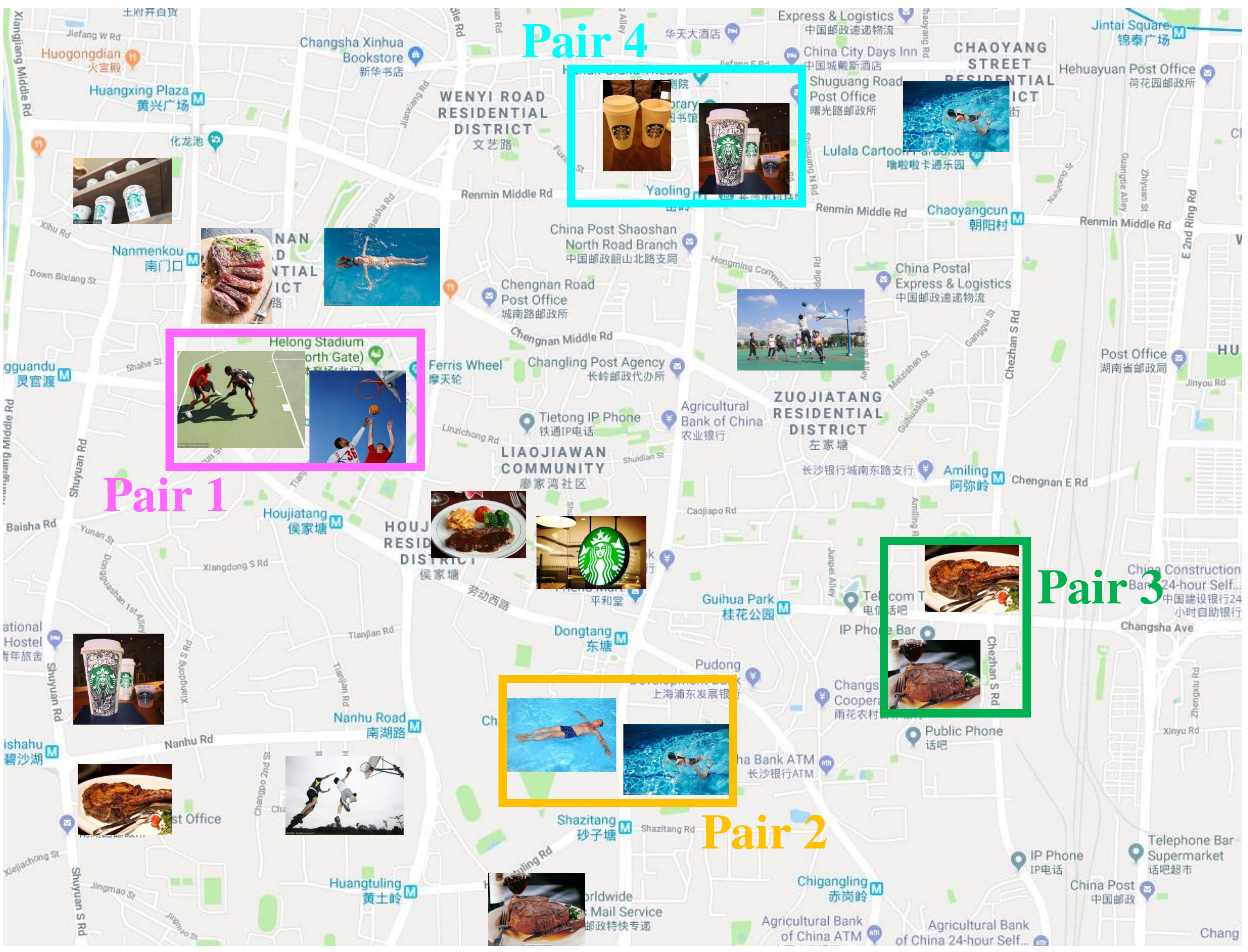}
   \captionsetup{justification=centering}
       \vspace{-0.2cm}
\caption{An example of spatial visual similarity join}
\label{fig:fig2}
\end{center}
\end{minipage}
\label{fig:k}
\end{figure*}

\noindent\textbf{Motivation}. Spatial textual search problem has become a hot spot in the area of spatial database and information retrieval due to the wide application of mobile devices and location-based services. Many spatial indexing techniques have been developed like R-Tree~\cite{DBLP:conf/sigmod/Guttman84}, R$^*$-Tree~\cite{DBLP:conf/sigmod/BeckmannKSS90}, IL-quadtree~\cite{DBLP:journals/tkde/ZhangZZL16}, KR$^*$-Tree~\cite{DBLP:conf/ssdbm/HariharanHLM07}, IR$^2$-Tree~\cite{DBLP:conf/icde/FelipeHR08} etc. Deng et al.~\cite{DBLP:journals/tkde/DengLLZ15} studied a generic
version of closest keywords search called best keyword Cover. Cao et al.~\cite{DBLP:conf/sigmod/CaoCJO11} proposed the problem of collective spatial keyword querying, Fan et al.~\cite{Fan2012Seal} studied the problem of spatio-textual similarity search for regions of interests query. However, these researches just only consider the textual data such as keywords and spatial information, they do not take into account other multimedia data mentioned above, like images. One of the important problem of spatial keyword search is named spatial textual similarity join, which is studied by many researches. It is to find out the spatial textual object pair which are similar in both aspects of geo-location and textual content. However, there is no work pay attention to multimedia data with geo-tags for this task. In this paper, we aim to investigate a novel paradigm named spatial visual similarity join and develop a efficient solution to overcome the challenge of geo-multimedia query. We in the first time propose a novel geo-multimedia query called region of visual interest query. Figure.~\ref{fig:fig2} is an simple but intuitive example to describe this problem.

\begin{example}
\label{ex:example1}
As illustrated in Fig.~\ref{fig:fig2}, the spatial visual similarity join can be applied in friends recommendation services on an online social networking platform. According to the geo-images posted by the users, the social networking system can find the similar geo-images in both aspects of geo-location and visual content. It's easy to understand that two people may make friend if they have the same hobbies and their position is very close. As illustrated in Fig.~\ref{fig:fig2}, there are four similar geo-image pair are search out. For pair 1, two users who took the photo about basketball at two close places are very likely to become good friends.
\end{example}

To our best knowledge, we are the first to propose the problem of spatial visual similarity join. To solve this problem effectively and efficiently, we present the definition of spatial visual similarity join at the first time and the relevant notions. Besides, we introduce how to measure geographical similarity and visual similarity to find the similar geo-image pairs. A baseline named SVS-JOIN$_B$ inspired by the techniques used in textual similarity join is introduced. Based on it, we propose an extension of SVS-JOIN$_B$ called SVS-JOIN$_G$ which uses spatial grid partition strategy to improve the efficiency. In order to furher improve the search performance, we develop a novel method named SVS-JOIN$_Q$, which is based on quadtree technique to partition the spatial region of input. A global inverted index is applied to enhance the search efficiency.

\noindent\textbf{Contributions}. Our main contributions can be summarized as follows:
\begin{itemize}
\item To the best of our knowledge, we are the first to study the problem of spatial visual similarity join. Firstly we propose the definition of geo-image and spatial visual similarity join and relevant notions. The visual similarity function and geographical similarity function are designed for similar geo-image pair search.
\item We introduce a baseline named SVS-JOIN$_B$ inspired by the techniques used for the problem of textual similarity join. An extension of SVS-JOIN$_B$ called SVS-JOIN$_G$ is developed, which has higher efficiency.
\item To further improve the searching performance , we present a novel method named SVS-JOIN$_Q$ based on quadtree partition technique and a global inverted index.
\item We have conducted extensive experiments on real geo-image dataset. Experimental results demonstrate that our solution has really high performance.
\end{itemize}

\noindent\textbf{Roadmap.} In the remainder of this paper, In Section~\ref{relwork} we present the related works about content-based image retrieval, spatial textual search and set similarity joins, which are related to our work. In Section~\ref{preliminaries} we propose the definition of spatial visual similarity join and related conceptions. We introduce a baseline and an extension named SVS-JOIN$_B$ and SVS-JOIN$_G$ respectively in Section~\ref{base}. In Section~\ref{globalindex}, we propose a novel technique named SVS-JOIN$_Q$ which utilize quadtree technique and a global inverted indexing structure to solve the spatial visual similarity join efficiently. In Section~\ref{perform} we present the experiment results. Finally, we conclude the paper in Section~\ref{con}.

\section{Related Work}
\label{relwork}

In this section, we introduce the previous studies of content-based image retrieval, spatial textual search, spatial queries on road networks and set similarity joins, which are relevant to this work. To the best of our knowledge, no priori work on this problem.

\noindent\textbf{Content-Based Image Retrieval.} As one of the most important problems, content-based image retrieval (CBIR for short) has gained much attention of many researchers due to benefit it provides for various multimedia analysis tasks~\cite{TC2018,DBLP:journals/pr/WuWLG18,DBLP:journals/ivc/WuW17}. Scale-Invariant Features Transform (SIFT for short) is one of the classical methods for visual feature extraction, which is proposed by Lowe~\cite{DBLP:conf/iccv/Lowe99}. It transforms an image into a collection of local feature vectors. These features are invariant to translation, scaling, rotation, and partially invariant to illumination changes. In another work of Lowe~\cite{DBLP:journals/ijcv/Lowe04}, Four main stages, i.e., (I) scale-space extrema detection, (II) keypoint localization, (III) orientation assignment, and (IV) keypoint descriptor, to generate the image features are proposed. Bag-of-Visual-Words (BoVW for short)~\cite{DBLP:conf/iccv/SivicZ03} is a traditional image representation model, which is to improve markedly the performance of image feature matching. This type of model was initially used to model texts and called Bag-of-Word (BoW for short). Specifically, for a textual document, this technique treats it as a collection of words and ignores the words order, syntax, etc. In this collection, the appearance of each word is independent. For image retrieval problem, it generates visual words by utilizing $k$-means method to cluster SIFT features. In recent years, lots of works have been proposed using SIFT and BoVW to overcome the challenges. For example, Mortensen et al.~\cite{DBLP:conf/cvpr/MortensenDS05} proposed a feature descriptor which augments SIFT with a global context vector that adds curvilinear shape information from a much larger neighborhood. This technique can improve the accuracy of image feature matching. Ke et al.~\cite{DBLP:conf/cvpr/KeS04} proposed a descriptor based on SIFT to encode the salient aspects of the image gradient in the neighborhood of feature point. Rather than using smoothed weighted histograms of SIFT, their method utilized Principal Components Analysis (PCA for short) to the normalized gradient patch. Su et al.~\cite{Su2017MBR} presented horizontal or vertical mirror reflection invariant binary descriptor named MBR-SIFT to solve the problem of image matching. In order to enhance the performance of matching, a fast matching algorithm is developed, which includes a coarse-to-fine two-step matching strategy in addition to two similarity measures. To gain sufficient distinctiveness and
robustness in the task of image feature matching, Li et al.~\cite{DBLP:journals/prl/LiM09} designed a novel framework based on SIFT for feature descriptor by integrating color and global information . Liao et al.~\cite{DBLP:journals/prl/LiaoLH13} presents an improvement to the SIFT descriptor for image matching and retrieval, which includes normalizing elliptical neighboring region, transforming to affine scale-space, improving the SIFT descriptor with polar histogram orientation bin, and integrating the mirror reflection invariant. Zhu et al~\cite{Zhu2013Image} proposed an image registration algorithm called BP-SIFT by using belief propagation, which have significant improvement for the problem of keypoint matching.

Originated from text retrieval and mining, BoVW is an important visual representation methods in multimedia retrieval and computer vision~\cite{DBLP:journals/tnn/WangZWLZ17,DBLP:journals/corr/abs-1804-11013,LINYANGARXIV,DBLP:conf/mm/WuWS13}. Escalante et al.~\cite{DBLP:journals/nca/EscalantePEBMM17} presented an evolutionary algorithm to implement an automatically learning weighting schemes of this model for computer vision tasks.In order to improve the performance of construction of visual words dictionary in the large image database, Dimitrovski et al.~\cite{DBLP:journals/isci/DimitrovskiKLD16} propose to use predictive clustering trees(PCTs) to improve the BoVW image retrieval, which can be constructed and executed efficiently and have good predictive performance. For the task of multi-script document retrieval, Mandal et al.~\cite{DBLP:journals/corr/abs-1807-06772} proposed a patch-based framework by using SIFT descriptor and bag-of-visual-words model to improve the performance of handwritten signature detection. Santos et al.~\cite{DBLP:journals/mta/SantosMST17} proposed a novel method based on S-BoVW paradigm that considers information of texture to generate textual signatures of image blocks. Moreover, they presented a strategy which represents image blocks with words which are generated based on both color as well as texture information. For the task of Medical image retrieval, Zhang et al.~\cite{DBLP:journals/ijon/ZhangSCHLPKFFC16} proposed a novel medical image retrieval approach named PD-LST retrieval, which is based on BoVW to identify discriminative characteristics between different medical images with Pruned Dictionary by using Latent Semantic Topic description. By utilizing BoVW representation, Karakasis et al.~\cite{Karakasis2015Image} proposed a novel framework for the task of image retrieval, which uses affine image moment invariants as descriptors of local image areas.

It is no doubt that these solutions improve the performance of image retrieval and visual feature matching significantly based on SIFT and BoVW model. However, these works cannot solve the problem of geo-multimedia data retrieval as they have no effective processing for geographical distance measurement.

\noindent\textbf{Spatial Textual Search.} Due to the collection and storage of large scale of spatial textual data, there has been increasing interest on spatial textual search problem in the community of spatial database. Spatial textual search~\cite{DBLP:conf/er/CaoCCJQSWY12,DBLP:journals/tkde/ZhangZZL16} aims to find out textual objects or documents with geo-tags by textual similarity and geographical proximity. For top-$k$ spatial keyword queries, Rocha-Junior et al.~\cite{DBLP:conf/ssd/RochaGJN11} propose a novel spatial index named Spatial Inverted Index (S2I for short) to enhance the efficiency of search. This technique maps each term in the vocabulary into a distinct aR-tree or block that stores all objects with the given term. Li et al.~\cite{DBLP:journals/tkde/LiLZLLW11} proposed an efficient indexing structure named IR-tree, which indexes both the textual and spatial contents of documents that enables spatial pruning and textual filtering to be performed at the same time. Furthermore, they developed a top-$k$ document search algorithm. Zhang et al~\cite{DBLP:conf/edbt/ZhangTT13} presented a scalable integrated inverted index called I$^3$ which uses the Quadtree structure to hierarchically partition the data space into cells. In order to improve the efficiency of retrieval, they designed a novel storage mechanism and preserve additional summary information to facilitate pruning, which outperform IR-tree and S2I in the aspects of construction time, index storage cost, updating speed and scalability. Zhang et al.~\cite{DBLP:conf/icde/ZhangZZL13} proposed a novel index structure named inverted linear quadtree (IL-Quadtree for short) which is based on the inverted index and the linear quadtree. Moreover, they designed a novel algorithm to improve the performance of query. Li et al.~\cite{Li2012Keyword} presented a novel spatial textual indexing technique named BR-tree to solve the problem of keyword-based $k$-nearest neighbor (K$^2$N$^2$ for short) queries, which utilizes R-tree to maintain the spatial information of objects and use the B-tree to main the terms in the objects. Fan et al.~\cite{Fan2012Seal} studied a novel search problem named spatio-textual similarity search which is to find the similar ROIs by considering spatial overlap and textual similarity. To improve the system performance, they proposed grid-based signatures and threshold-aware pruning techniques. Zhang et al~\cite{DBLP:conf/sigir/ZhangCT14} proposed a novel method which is based on modeling the spatial keyword search problem as a top-$k$ aggregation problem. They developed a rank-aware CA algorithm which works well on inverted lists sorted by textual relevance and spatial curving order. Lin et al.~\cite{DBLP:journals/tkde/LinXH15} propose a novel spatial textual query paradigm called reverse keyword search for spatio-textual top-$k$ queries (RSTQ for short). They developed a novel hybrid indexing structure named KcR-tree to store and summarize the spatial and textual information of objects and proposed three query optimization techniques, i.e., KcR$^*$-tree, lazy upper-bound updating, and keyword set filtering to improve the performance of search. For the problem of continuous spatial-keyword queries over streaming data, Wang et al.~\cite{DBLP:conf/icde/WangZZLW15} proposed a highly efficient indexing technique named AP-Tree which adaptively groups registered queries utilizing keywords and spatial partitions based on a cost model and indexes ordered keyword combinations. Besides, they devised an index construction algorithm that seamlessly and effectively combines keyword and spatial partitions. Zheng et al.~\cite{DBLP:conf/icde/ZhengSZSXLZ15} investigated an other type of spatial textual query, namely interactive top-$k$ spatial keyword queries. They introduced a three-phase solution: the first phase is to quickly narrows the search space from the entire database. The second phase is called interactive phase, which is to develop several strategies to select a subset of candidates and present them to users at each round. The third phase is to terminate the interaction automatically. Zhang et al.~\cite{DBLP:conf/icde/ZhangCMTK09} introduced $m$-closest keywords ($m$CK for short) query which aims to search out the spatially closest tuples which match $m$ user-specified keywords. To solve this problem more efficiently, they developed a novel spatial index called the bR$^*$-tree extending from the R$^*$-tree and proposed a priori-based search strategies to effectively reduce the search space. Guo et al.~\cite{DBLP:conf/sigmod/GuoCC15} proposed another solution to solve the $m$CK search problem. They devised a novel greedy algorithm named $SKEC$ that has an approximation ratio of 2 and in addition, they developed another two approximation algorithms called $SKECa$ and $SKECa+$ respectively to improve the efficiency. Besides, a novel exact algorithm utilizing $SKECa+$ is introduced to reduce the search space considerably.

The researches aforementioned focused on textual search based on Euclidean distance, they do not consider the queries on road network which is a is a more realistic situation. On the other hand, these solutions cannot be applied in the problem of multimedia retrieval.

\noindent\textbf{Spatial Queries on Road Networks} In order to better simulate real life situations, more and more researchers are starting to pay attention to the problem of spatial queries on road networks. For example, Lee et al.~\cite{DBLP:conf/edbt/LeeLZ09} presented a general framework named ROAD to evaluate Location-Dependent Spatial Queries (LDSQ)s on road networks. Li et al.~\cite{DBLP:conf/dasfaa/LiGZ14} studied the problem of range-constrained spatial keyword query on road networks and proposed three approaches, i.e., expansion-based approach (EA), Euclidean heuristic approach (EHA) and Rnet Hierarchy based approach (RHA). Zhang et al.~\cite{DBLP:conf/edbt/ZhangZZLCW14} presented a signature-based inverted indexing technique and an efficient diversified spatial keyword search algorithm to solve the problem of diversified spatial keyword search on road networks. Rocha-Junior et al.~\cite{DBLP:conf/edbt/Rocha-JuniorN12} at first time studied the problem of top-$k$ spatial keyword queries on road networks. They devised novel indexing structures and algorithms to deal with this problem efficiently. For the problem of collective spatial keyword query (CSKQ for short) processing on road networks. Gao et al.~\cite{DBLP:journals/tits/GaoZZC16} proved that it is a NP-complete at first time and designed two approximate algorithms with provable approximation bounds and one exact algorithm. For $k$ nearest neighbor ($k$NN) query, Zhong et al.~\cite{DBLP:journals/tkde/ZhongLTZG15} presented a novel indexing structure named G-tree which is a height-balanced and scalable index inspired by R-tree, and then developed efficient search algorithms for SPSP, $k$NN and k$^2$N$^2$ query. Abeywickrama et al.~\cite{Abeywickrama2016k} studied the problem of $k$ nearest neighbor query on road networks and presented efficient implementations of five of the most notable methods, i.e., IER, INE, Distance Browsing, ROAD and G-tree. For aggregate nearest neighbor (ANN for short) queries problem, Sun et al.~\cite{Sun2015On} proposed effective pruning strategies for sum aggregate function and max aggregate function, and developed an efficient NVD-based algorithm. In another work of Gao~\cite{Gao2015Efficient}, reverse top-k boolean spatial keyword queries on road networks was investigated. They formalized the RkBSK query and presented filter-and-refinement framework based algorithms to solve this problem.

\noindent\textbf{Set Similarity Joins}
In recent years, lots of researchers paid attentions on the problem of spatial textual similarity join. A spatial similarity join of two spatial databases aims to search out pairs of objects that are simultaneously similar in both the aspects of textual and spatial. Ballesteros et al~\cite{Ballesteros2011SpSJoin} proposed an algorithm based on MapReduce parallel programming model to solve this problem on large-scale spatial databases.

\section{Preliminaries}
\label{preliminaries}
In this section, we propose the definition of region of visual interests (RoVI for short) at the first time, then present the notion of region of visual interests query (RoVIQ for short) and the similarity measurement. Besides, we review the techniques of image retrieval which is the base of our work. Table~\ref{tab:notation} summarizes the notations frequently used throughout this paper to facilitate the discussion.

\begin{table}
	\centering
    \small
	\begin{tabular}{|p{0.15\columnwidth}| p{0.78\columnwidth} |}
		\hline
		\textbf{Notation} & \textbf{Definition} \\ \hline\hline
		~$\mathcal{D}_I$                                 & A given database of geo-images                \\ \hline
        ~$|\mathcal{D}_I|$                               & The number of geo-images in $\mathcal{D}_I$               \\ \hline
        ~$I_i$                                           & The $i$-th geo-tagged images                 \\ \hline
        ~$I_i.G$                                         & The geographical information component of $I_i$      \\ \hline
        ~$I_i.V$                                         & The visual component of $I_i$                 \\ \hline
        ~$X$                                             & A longitude                \\ \hline
        ~$Y$                                             & A latitude        \\ \hline
	  	~$v_i$                                           & A visual word         \\ \hline
        ~$\mathcal{R}_I$                                 & A dataset of geo-images         \\ \hline
        ~$I^r_i$                                         & The $i$-th geo-tagged images in dataset $\mathcal{R}$   \\ \hline
        ~$w(v)$                                          & The weight of visual word $v$          \\ \hline
        ~$\Gamma_G$                                      & The geographical similarity threshold          \\ \hline
        ~$\Gamma_V$                                      & The visual similarity threshold          \\ \hline
        ~$\mathcal{P}$                                   & A result set        \\ \hline
        ~$GeoSim(I_i,I_j)$                               & The geographical similarity between $I_i$ and $I_j$     \\ \hline
		~$VisSim(I_I,i_J)$                               & The visual similarity between $I_i$ and $I_j$     \\ \hline
        ~$EucDis(I_I,i_J)$                               & The Euclidean distance between $I_i$ and $I_j$     \\ \hline
        ~$MaxDis(\mathcal{R},\mathcal{S})$               & The maximum Euclidean distance between any two geo-iamges from $\mathcal{R}$ and $\mathcal{S}$             \\ \hline
        ~$\mathcal{W}$                                   & A global word set        \\ \hline
        ~$\mathcal{L}$                                   & A inverted list        \\ \hline
        ~$\Omega_o[\hat{o}]$                             & The number of word overlap of $o$ with $\hat{o}$      \\ \hline
        ~$\Phi$                                          & A global word ordering          \\ \hline
        ~$Pf(I.V)$                                       & The prefix of $I.V$          \\ \hline
        ~$C_i$                                           & A cell with id $i$          \\ \hline
        ~$N_i$                                           & A node of quadtree with id $i$          \\ \hline
        ~$\mathcal{I}_G$                                 & A inverted index set          \\ \hline
	\end{tabular}
    \caption{The summary of notations} \label{tab:notation}	
\end{table}

\subsection{Problem Definition}
\begin{definition}[\textbf{Geo-Image}] \label{def:geo-image}
Let $\mathcal{D}_I = \{I_1,I_2,...,I_{|\mathcal{D}_I|}\}$ be an geo-image dataset, $|\mathcal{D}_I|$ denotes the size of $\mathcal{D}_I$. A geo-image $I_i \in \mathcal{D}_I$ is defined as a tuple $I_i=<I_i.G,I_i.V>$, where $I_i.G$ is the geographical information component which is generated from the geo-tag of this image. More specifically, it consists of longitude $X$ and latitude $Y$, i.e., $I_i.G = (X,Y)$. Another part, $I_i.V$, is the visual information component which is a visual word set $I_i.V = \{v_1,v_2,...,v_n\}$ modeled by SIFT and BoVW. $i$ is the id of this geo-image.
\end{definition}

Consider two geo-image datasets $\mathcal{R}_I = \{I^r_1,I^r_2,...,I^r_{|\mathcal{R}_I|}\}$ and $\mathcal{S}_I=\{I^s_1,I^s_2,...,I^s_{|\mathcal{S}_I|}\}$, similar to spatial textual similarity join, a spatial visual similarity join aims to search out all pairs of geo-images from $\mathcal{R}_I$ and $\mathcal{S}_I$ respectively, which are similar enough in both aspects of geo-location and visual content. We introduce two thresholds, i.e., geographical similarity threshold and visual similarity threshold to measure these two similarity. Specifically, for each pair, both of the geographical similarity and visual similarity of this two geo-images are less than geographical similarity threshold and visual similarity threshold. In order to clarify our work more clearly, we proposed the definition of spatial visual similarity join as follows.

\begin{definition}[\textbf{Spatial Visual Similarity Join (SVS-JOIN)}] \label{def:SVS-Join}
Given two geo-image datasets $\mathcal{R}_I = \{I^r_1,I^r_2,...,I^r_{|\mathcal{R}_I|}\}$ and $\mathcal{S}_I=\{I^s_1,I^s_2,...,I^s_{|\mathcal{S}_I|}\}$, geographical similarity threshold $\Gamma_G$ and visual similarity threshold $\Gamma_V$. A spatial visual similarity join denoted as $\mbox{SVS-JOIN}(\mathcal{R},\mathcal{S},\Gamma_G,\Gamma_V)$ returns a set of geo-image pairs $\mathcal{P} \subseteq \mathcal{R} \times \mathcal{S}$, in which each pair contains two highly similar geo-images in both aspect of geo-location and visual content, i.e.,
\begin{equation*}
  \mathcal{P}=\{(I^r_i,I^s_j)|GeoSim(I^r_i,I^s_j) \leq \Gamma_G, VisSim(I^r_i,I^s_j) \geq \Gamma_V, \forall I^r_i \in \mathcal{R}_I, I^s_j \in \mathcal{S}_I\}
\end{equation*}
where $GeoSim(I^r_i,I^s_j)$ and $VisSim(I^r_i,I^s_j)$ are the geographical similarity function and visual similarity function respectively.
\end{definition}

To measure these two similarities quantitatively, in this work we utilize Euclidean distance measurement and Jaccard distance measurement to construct our functions, which are described formally as follows.

\begin{definition}[\textbf{geographical similarity function}] \label{def:geo-sim-func}
Given two geo-image datasets $\mathcal{R}_I = \{I^r_1,I^r_2,...,I^r_{|\mathcal{R}_I|}\}$ and $\mathcal{S}_I=\{I^s_1,I^s_2,...,I^s_{|\mathcal{S}_I|}\}$, $\forall I^r_i \in \mathcal{R}_I, I^s_j \in \mathcal{S}$, the geographical similarity between $I^r_i$ and $I^s_j$ is measured by the following similarity function:
\begin{equation}\label{euq:geo-sim-func}
  GeoSim(I^r_i,I^s_j)=1-\frac{EucDis(I^r_i,I^s_j)}{MaxDis(\mathcal{R},\mathcal{S})}
\end{equation}
where $EucDst(I^r_i,I^s_j)$ is the Euclidean distance between $I^r_i$ and $I^s_j$, which is measured by the following function:
\begin{equation}\label{equ:EcuDis}
  EucDis(I^r_i,I^s_j) = \sqrt{(I^r_i.G.X-I^s_j.G.X)^2+(I^r_i.G.Y-I^s_j.G.Y)^2}
\end{equation}
the function $MaxDis(\mathcal{R},\mathcal{S})$ is to return the maximum Euclidean distance between any two geo-images from $\mathcal{R}$ and $\mathcal{S}$ respectively, which is described in formal as follows:
\begin{equation}\label{equ:MaxDis}
  MaxDis(\mathcal{R},\mathcal{S}) = max(\{EucDis(I^r_i,I^s_j)|I^r_i \in \mathcal{R}, I^s_j \in \mathcal{S}\})
\end{equation}
where the function $max(\cdot)$ is to return the maximum element from a set.
\end{definition}

\begin{definition}[\textbf{visual similarity function}] \label{def:vis-sim-func}
Given two geo-image datasets $\mathcal{R}_I = \{I^r_1,I^r_2,...,I^r_{|\mathcal{R}_I|}\}$ and $\mathcal{S}_I=\{I^s_1,I^s_2,...,I^s_{|\mathcal{S}_I|}\}$, $\forall I^r_i \in \mathcal{R}_I, I^s_j \in \mathcal{S}$, the visual similarity between $I^r_i$ and $I^s_j$ is measured by the following similarity function:
\begin{equation}\label{equ:vis-sim-func}
  VisSim(I^r_i,I^s_j)=1-\frac{\sum_{v \in I^r_i.V \cap I^s_j.V}^{}w(v)}{\sum_{v \in I^r_i.V \cup I^s_j.V}^{}w(v)}
\end{equation}
where $w(v)$ represents the weight of the visual word $v$. In this work, we measure the weight of visual word by inverse document frequency $idf$.
\end{definition}

\noindent\textbf{Assumption}. For ease of discussion, in this work we assume that $\mathcal{R}=\mathcal{S}$. Our approach can be applied well in the case of $\mathcal{R} \neq \mathcal{S}$. Therefore, for a geo-image dataset $\mathcal{R}$, we denote a spatial visual similarity join as $\mbox{SVS-JOIN}(\mathcal{R},\Gamma_G,\Gamma_V)$.

\begin{figure*}
\newskip\subfigtoppskip \subfigtopskip = -0.1cm
\begin{minipage}[b]{0.99\linewidth}
\begin{center}
     \includegraphics[width=1\linewidth]{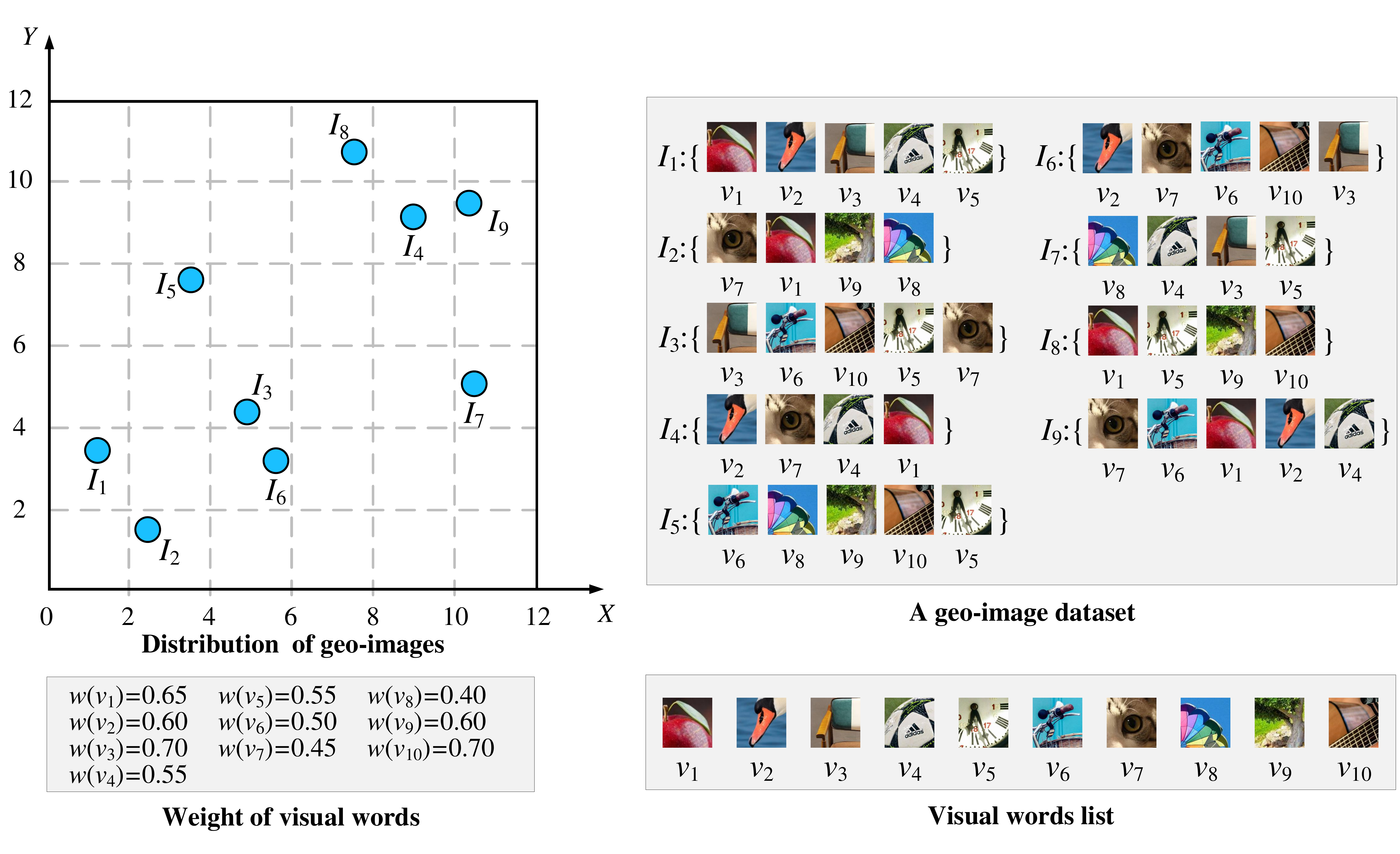}
   \captionsetup{justification=centering}
       \vspace{-0.2cm}
\caption{An example of geo-images and spatial visual similarity join}
\label{fig:fig-example-geo-images}
\end{center}
\end{minipage}
\label{fig:k}
\end{figure*}

\begin{example} \label{ex:geo-images}
Consider a geo-image dataset $\mathcal{R} = \{I_1,I_2,...,I_9\}$, shown in Fig.~\ref{fig:fig-example-geo-images}. We set $\Gamma_G = 0.3$ and $\Gamma_V=0.4$, the $\mbox{SVS-JOIN}(\mathcal{R},0.3,0.4)$ return the set $\mathcal{P}=\{(I_3,I_6),(I_4,I_9)\}$.
\end{example}

\subsection{Visual Feature Extraction and Image Representation}

\noindent\textbf{Visual Feature Extraction}. In this work, we utilize SIFT~\cite{DBLP:journals/ijcv/Lowe04} which is a important traditional technique for the task of visual feature extraction. SIFT aims to transforms an image into a large set of local feature vectors. This feature vectors are  invariant to image translation, scaling, and rotation, and partially invariant to illumination changes and affine or 3D projection. There are four main phases of visual feature extraction by SIFT:
\begin{itemize}
\item \textbf{Scale-space extrema detection}. The first phase is called scale-space extrema detection. This method searches all the images in scale space, which is to identify potential points of interest that are invariant to scale and orientation by utilizing difference-of-Gaussian (DoG) function shown as follows:
    \begin{equation}\label{equ:dog}
       D(x,y,\sigma)=(G(x,y,k\sigma)-G(x,y,\sigma))*I(x,y)
    \end{equation}
    where $I(x,y)$ represents an image and the operator $*$ represents convolution operation. $G(x,y,\sigma)$ is the Gaussian kernel function shown as follows:
    \begin{equation}\label{equ:guassian}
      G(x,y,\sigma) = \frac{1}{2 \pi \sigma^2}exp(-(x^2+y^2)/2 \sigma^2)
    \end{equation}
\item \textbf{Keypoint localization}. The second phrase is named keypoint localization, which is to select and localize the keypoints according to their stability. At each candidate location, a fine fitting model is used to determine the location and scale. In this process, there are two type of keypoints should be rejected: (I) low contrast feature points and (II) unstable edge response points. For the first type of keypoint $X = (x,y,\sigma)^T$, the offset is defined as $\Delta X$, then the method uses the Taylor expansion of $D(x,y,\sigma)$,
    \begin{equation}\label{equ:taylor-expansion-D}
      D(X) = D + \frac{\partial D^T}{\partial X}\Delta X + \frac{1}{2}\Delta X^2\frac{\partial^2 D}{\partial X^2}\Delta X
    \end{equation}
    as $X$ is the extremum of $D(X)$ determined by taking the derivative of this function with respect to $X$ and setting it to zero,
    \begin{equation}\label{equ:derivative}
      \Delta X = -\frac{\partial^2 D^{-1}}{\partial X^2}\frac{\partial D}{\partial X}
    \end{equation}
    thus we can use the function value at the extremum to reject unstable extrema with low contrast, shown as follows,
    \begin{equation}\label{equ:extremum}
      D(\Delta X = D + \frac{1}{2}\frac{\partial D^T}{\partial X}\Delta X
    \end{equation}
    let $T$ be the threshold of contrast, if $|D(\Delta X)| < T$, then the keypoint is rejected.
    For type (II) keypoints, which are unstable edge response points, this method calculates the eigenvalue of $2 \times 2$ Hessian matrix $\textbf{H}$, which is in direct proportion to the principal curvature of $DoG$ function $D(X$) of candidate feature points,
    \begin{equation}\label{equ:Hessian-matrix}
     \textbf{H}=\left[
                \begin{matrix}
                   D_{xx} & D_{xy} \\
                   D_{yx} & D_{yy} \\
                \end{matrix}
                \right]
    \end{equation}
    Then, we can compute the trace and determinant of $\textbf{H}$ as follows, we set $\alpha = \lambda_{max}$ as the maximum eigenvalue of $\textbf{H}$, $\beta = \lambda_{min}$ as the minimum eigenvalue of $\textbf{H}$, then
    \begin{equation}\label{equ:trace}
       Tr(\textbf{H}) = D_{xx}+D_{yy} = \alpha+\beta
    \end{equation}
    \begin{equation}\label{equ:determinant}
       Det(\textbf{H})=D_{xx}D_{yy}-(D_{xy})^2=\alpha \beta
    \end{equation}
    let $\gamma$ be the ratio of maximum eigenvalue to minimum eigenvalue, then
    \begin{equation}\label{ratio}
      \frac{Tr(\textbf{H})^2}{Det(\textbf{H})} = \frac{(r+1)^2}{r}
    \end{equation}
    Thus, in order to detect whether the principal curvature is below a threshold value $\tau_{\gamma}$, we just to check the following inequality
    \begin{equation}\label{check-threshold}
      \frac{Tr(\textbf{H})^2}{Det(\textbf{H})} \geq \frac{(\tau_{\gamma}+1)^2}{\tau_{\gamma}}
    \end{equation}
    If this inequality is established, the keypoint is rejected.
\item \textbf{Orientation assignment}. In the orientation assignment phrase, according to the local gradient direction of the image, each keypoint is assigned one or more directions, and all subsequent operations transform the direction, scale and position of the keypoints to provide invariance of features to these transformations. According to the gradient distribution characteristics of neighborhood pixels of keypoints to determine its direction parameters, and then using the gradient histogram of the image to obtain the stable direction of local structure of key points. The scale image of feature points can be calculated by
    \begin{equation}\label{equ:scale-image}
      L(x,y)=G(x,y,\sigma)*I(x,y)
    \end{equation}
    Thus, for each image $L(x,y)$, the gradient magnitude $m(x,y)$ and the orientation $\theta(x,y)$ can be calculated respectively by the following equations:
    \begin{equation}\label{gradient-magnitude}
      m(x,y)=\sqrt{(L(x+1,y)-L(x-1,y))^2+(L(x,y+1)-L(x,y-1))^2}
    \end{equation}
    \begin{equation}\label{orientation}
      \theta(x,y)=arctan\frac{L(x,y+1)-L(x,y-1)}{L(x+1,y)-L(x-1,y)}
    \end{equation}
    After calculating the gradient direction, the orientation histogram can be generated from the gradient orientations of sample points within a region around the keypoint, which is used to calculate the gradient direction and amplitude of the pixels in the neighborhood of the feature points. The peaks of histogram are the dominant directions of keypoints.
\item \textbf{Keypoint descriptor}. In the last phase, the local gradients of the image are measured around each feature point at selected scales. And these gradients are transformed into a representation which allows for significant local shape distortion and illumination transformation.
\end{itemize}

\section{The Baseline for Spatial Visual Similarity Joins}
\label{base}

In this section, we propose the baseline for the problem of spatial visual similarity joins. Firstly, we introduce the state-of-the-art algorithm named PPJOIN~\cite{DBLP:journals/tods/XiaoWLYW11} for textual similarity joins, which is utilized in our baseline. Then we present our baseline named SVS-JOIN$_B$ in detail.

\subsection{The Method for Textual Similarity Joins}
\noindent\textbf{Inverted Index Based Method}. The traditional way to solve textual similarity joins efficiently is to construct a inverted index for the target object dataset $\mathcal{R}$, which associates each word in the global word set $\mathcal{W}$ built beforehand to an objects inverted list $\mathcal{L}$. For each object $o \in \mathcal{R}$, the inverted list $\mathcal{L}(w)$ of each word $w \in o.V$ is traversed, where $o.V$ represents the keywords set of $o$. Then we account the number of word overlap of $o$ with every object $\hat{o} \in \mathcal{L}(w)$ and save these numbers in a set $\Omega_o$. To facilitate exploration of this method, we denote $\Omega_o[\hat{o}]$ as the number of word overlap of $o$ with $\hat{o}$. Apparently, the candidate pair set can be generated from $\Omega_o$ directly. Then for all objects $\hat{o} \in \Omega_o$, if $\Omega_o[\hat{o}] > 0$ and $TxtSim(o,\hat{o}) \geq \theta_t$, then the pair $(o,\hat{o})$ can be one of the results. To describe this process in a formal way, textual similarity joins by this method is to return a result set $\mathcal{P}$,
\begin{equation}
  \mathcal{P} = \{(o,\hat{o})|o \in \mathcal{R}, \hat{o} \in \Omega_o, \Omega_o[\hat{o}] > 0 \ and \ TxtSim(o,\hat{o}) \geq \theta_t\}
\end{equation}
where $TxtSim(o,\hat{o})$ is the textual similarity function and $\theta_t$ is the textual similarity threshold.

\noindent\textbf{Prefix Filtering Principle}. When we use the inverted index based method, the inverted list $\mathcal{L}_w$ will be quite long if the word $w$ is very frequent in the dataset. This become a major challenge as a lot of candidate pairs have to be generated in this situation. In order to reduce the size of the candidate set, an efficient method called prefix filtering principle was devised by~\cite{DBLP:conf/icde/ChaudhuriGK06}. According to this technique, we generate a global word ordering $\Phi$ which sorts keywords by word frequency in reverse order, and then for all objects $o \in \mathcal{R}$, order the keywords in $o.V$ by $\Phi$. After the ordering, the prefix of $o.V$ is denoted as $Pf(o.V)$ and the length of it is denoted as $|Pf(o.V)|$, which is measured by the following equation:
\begin{equation}\label{length-of-prefix}
  |Pf(o.V)| = |o.V| - \lceil \theta_t |o.V| \rceil + 1
\end{equation}
where $|o.V|$ represents the number of keywords in $o.V$, $\theta_t$ is the textual similarity threshold. It is obvious that the length of prefix of an object is determined by the number of keywords contained by this object and the similarity threshold given in advance. According to this principle, we can get the following theorem:
\begin{theorem} \label{thm:prefix}
  Given two objects $o,\hat{o} \in \mathcal{R}$ and a textual similarity threshold $\theta_t$, if \ $TxtSim(\\o,\hat{o}) \geq \theta_t$, then $Pf(o.V) \cap Pf(\hat{o}.V) \neq \emptyset$.
\end{theorem}

Obviously, the basic idea of Theorem~\ref{thm:prefix} is that if the textual similarity between two objects is larger than a threshold, they should be share same keywords. Therefore, this theorem can be used to prune the candidate pair set effectively. Specifically, for each object $o$, we just only to search out the keywords contained in the prefix of $o$.

\noindent\textbf{The PPJOIN Algorithm}. PPJOIN is one of the efficient algorithm to solve the textual similarity joins problem, developed by Xiao et al.~\cite{DBLP:journals/tods/XiaoWLYW11}.

\begin{algorithm}
\begin{algorithmic}[1]
\footnotesize
\caption{\bf PPJOIN Algorithm}
\label{alg:ppjoin}

\INPUT $\mathcal{R}$ is an objects dataset sorted by a global ordering $\Phi$, $\theta_t$ is a textual similarity threshold.
\OUTPUT $\mathcal{P}$ is the result pairs set.

\FOR{each $w \in \mathcal{W}$}
    \STATE $\mathcal{L}_w \leftarrow \emptyset$;
\ENDFOR

\FOR{each $o \in \mathcal{R}$}
    \STATE $|Pf(o.V)|_S \leftarrow |o.V|-\lceil \theta_t |o.V| \rceil + 1$;
    \STATE $|Pf(o.V)|_I \leftarrow |o.V|-\lceil \frac{2\theta_t}{\theta_t+1} |o.V| \rceil + 1$;
    \FOR{$i=1$ to $|Pf(o.V)|_S$}
        \STATE $w \leftarrow$ the $i$-th keyword in $o.V$;
        \FOR{$e(\hat{o},i_{\hat{o}}) \in \mathcal{L}_w$ and $|\hat{o}.V| \geq \theta_t |o.V|$}
            \IF{$QualifyPosFilter(o,i_o,\hat{o},i_{\hat{o}}) \ \&\& \ QualifySufFilter(o,i_o,\hat{o},i_{\hat{o}})$}
                \STATE $\Omega_o[\hat{o}] \leftarrow \Omega_o[\hat{o}]+1$;
            \ELSE
                \STATE $\Omega_o[\hat{o}] \leftarrow -\infty$;
            \ENDIF
        \ENDFOR
        \IF{$i_o \leq |Pf(o.V)|_I$}
            \STATE $\mathcal{L}_w \leftarrow \mathcal{L} \cup \{(o,i_o)\}$
        \ENDIF
    \ENDFOR
    \STATE $Verify(o,\Omega_o,\mathcal{P})$;
\ENDFOR
\RETURN $\mathcal{P}$;
\end{algorithmic}
\end{algorithm}

Algorithm demonstrates the pseudo-code of the PPJOIN algorithm. The input of this algorithm is a textual similarity threshold $\theta_t$ and an objects dataset $\mathcal{R}$ which is sorted in ascending order of their size. At first it generates inverted list for each word in the global words set. Then for each object $o \in \mathcal{R}$, probe prefix length $|Pf(o.V)|_S$ and $|Pf(o.V)|_I$ are calculated. Then from the first position to the $|Pf(o.V)|_S$-th position, it scans the prefix of $o$ and get the word in the prefix, and generates candidate pair. After that, the filter condition $|\hat{o}.V| \geq \theta_t |o.V|$ is used by this algorithm to filters the candidate pairs. The positional and suffix filter are operated by calling two produces $QualifyPosFilter(o,i_o,\hat{o},i_{\hat{o}})$ and $QualifySufFilter(o,i_o,\hat{o},i_{\hat{o}})$. The overlap will be added if the pair can qualify these filters. At last, this algorithm generates the result set $\mathcal{P}$ by executing the procedure $Verify(o,\Omega_o,\mathcal{P})$.

\subsection{The Baselines for Spatial Visual Similarity Joins}
In this subsection, we introduce the our baseline approach. Inspiring by the prefix filtering principle and the PPJOIN algorithm, we proposed a baseline called SVS-JOIN$_B$ for the problem of spatial visual similarity joins. Different from the textual similarity joins, our method consider two aspects of information, i.e., geographical information and visual information. We set two thresholds $\Gamma_G$ and $\Gamma_V$ to deal with the measurement of geographical similarity and visual similarity. According to the definition of spatial visual similarity joins, we implement two procedures $GeoSim(I,\hat{I})$ and $VisSim(I,\hat{I})$ to calculate these two similarities.

\noindent\textbf{SVS-JOIN$_B$ Algorithm}. Algorithm~\ref{alg:svs-joins_b} demonstrates the computing process in the form of pesudo-code. The input is a geo-image dataset and two thresholds $\Gamma_G$ and $\Gamma_V$. Different from Algorithm~\ref{alg:ppjoin}, in Line 9, $GeoSim(I,\hat{I})$ is called as a geographical similarity filter to prune the geo-image pairs whose spatial distance between two images is not short enough. Like PPJOIN, the procedure $Verify(I,\Omega_I,\mathcal{P})$ generates the final results set from the candidate set based on $\Omega_I$.

\begin{algorithm}
\begin{algorithmic}[1]
\footnotesize
\caption{\bf SVS-JOIN$_B$ Algorithm}
\label{alg:svs-joins_b}

\INPUT $\mathcal{R}$ is a geo-image dataset sorted by a global ordering $\Phi$, $\Gamma_V$ is a textual similarity threshold, $\Gamma_G$ is a geographical similarity threshold.
\OUTPUT $\mathcal{P}$ is the result pairs set.

\FOR{each $w \in \mathcal{W}$}
    \STATE $\mathcal{L}_v \leftarrow \emptyset$;
\ENDFOR

\FOR{each $o \in \mathcal{R}$}
    \STATE $|Pf(I.V)|_S \leftarrow |I.V|-\lceil \Gamma_V |I.V| \rceil + 1$;
    \STATE $|Pf(I.V)|_I \leftarrow |I.V|-\lceil \frac{2\Gamma_V}{\Gamma_V+1} |I.V| \rceil + 1$;
    \FOR{$i=1$ to $|Pf(I.V)|_S$}
        \STATE $V \leftarrow$ the $i$-th visual word in $o.V$;
        \FOR{$e(\hat{I},i_{\hat{I}}) \in \mathcal{L}_v$ and $GeoSim(I,\hat{I}) \leq \Gamma_G$ and $|\hat{o}.V| \geq \Gamma_V |o.V|$}
            \IF{$QualifyPosFilter(I,i_I,\hat{I},i_{\hat{I}}) \ \&\& \ QualifySufFilter(I,i_I,\hat{I},i_{\hat{I}})$}
                \STATE $\Omega_I[\hat{I}] \leftarrow \Omega_I[\hat{I}]+1$;
            \ELSE
                \STATE $\Omega_I[\hat{I}] \leftarrow -\infty$;
            \ENDIF
        \ENDFOR
        \IF{$i_I \leq |Pf(I.V)|_I$}
            \STATE $\mathcal{L}_v \leftarrow \mathcal{L} \cup \{(I,i_I)\}$
        \ENDIF
    \ENDFOR
    \STATE $Verify(I,\Omega_I,\mathcal{P})$;
\ENDFOR
\RETURN $\mathcal{P}$;
\end{algorithmic}
\end{algorithm}

Although SVS-JOIN$_B$ algorithm can effectively deal with the problem of spatial visual similarity joins, we can still improve the efficiency significantly. It is easily to know that we just consider the geo-image $\hat{I}$ who satisfies the filter condition $GeoSim(I,\hat{I}) \leq \Gamma_G$. Unfortunately, this is the main limitation. In other words, SVS-JOIN$_B$ algorithm considers all the geo-images $\hat{I} \in \mathcal{L}_v$ for each visual word $v$ which is contained in $Pf(I.V)$. To overcome this challenge, in the next part we present a grid based spatial partition strategy and develop a more efficient baseline named SVS-JOIN$_G$ algorithm extending from SVS-JOIN$_B$.

\noindent\textbf{Spatial Grid}. For the task of spatial visual similarity joins, we propose a grid based spatial partition strategy named spatial grid to improve the performance of algorithm. This strategy is to model the two-dimensional spatial area of a geo-image dataset $\mathcal{R}$ as a grid, denoted as $G(\mathcal{R})$ contained several cells which equals to the geographical similarity threshold $\Gamma_G$ in each dimension. Thus, the area of each cell equals $\Gamma_G^2$. It is clear that the spatial grid is determined by a spatial visual similarity join with dataset $\mathcal{R}$ and threshold $\Gamma_G$. To put it in another way, for a given dataset $\mathcal{R}$, the grid $G(\mathcal{R})$ do not need to pre-compute. 

\begin{figure*}
\newskip\subfigtoppskip \subfigtopskip = -0.1cm
\begin{minipage}[b]{0.99\linewidth}
\begin{center}
     \subfigure[{the use of spatial grid}]{
     \includegraphics[width=0.48\linewidth]{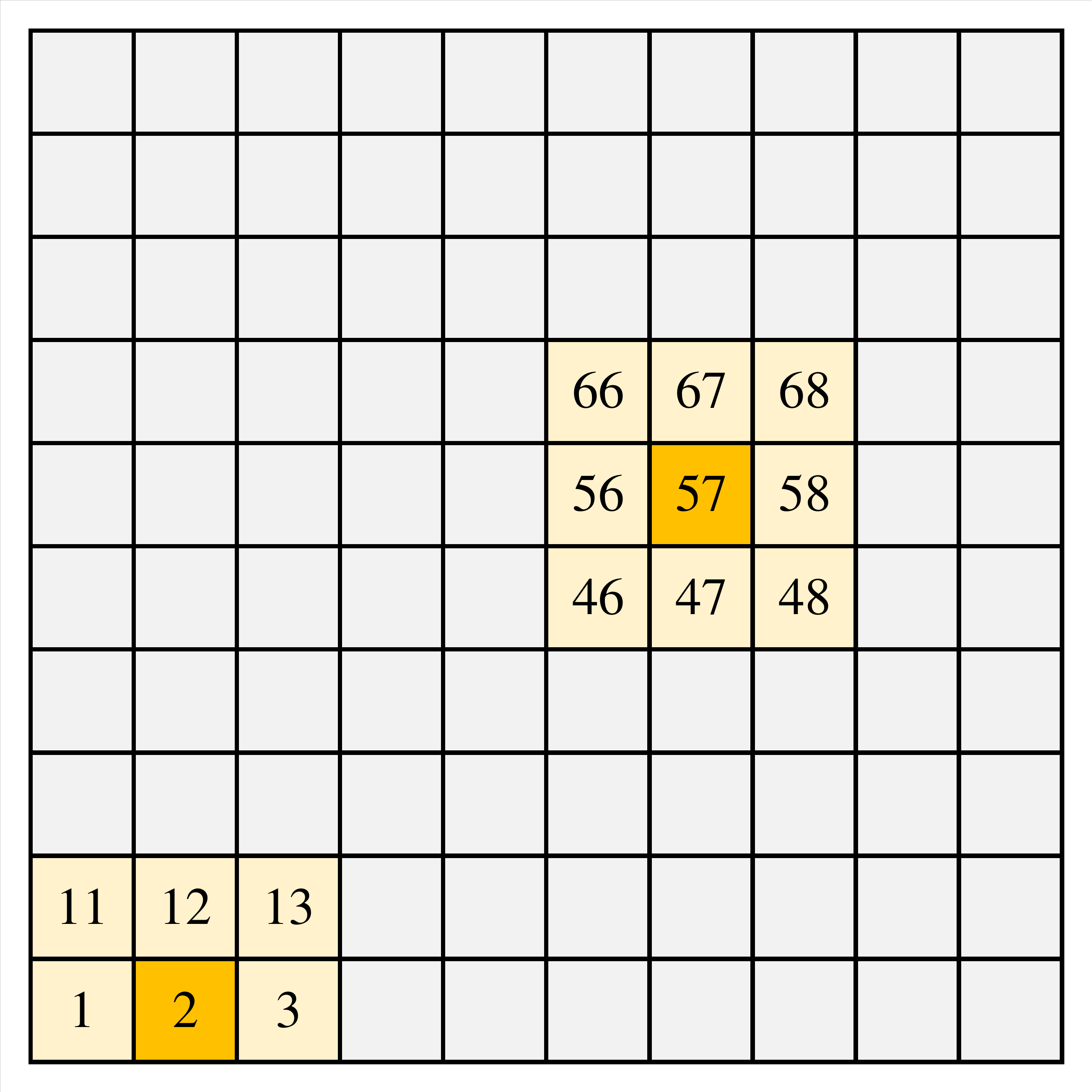}
     }
     \subfigure[{}]{
     \includegraphics[width=0.48\linewidth]{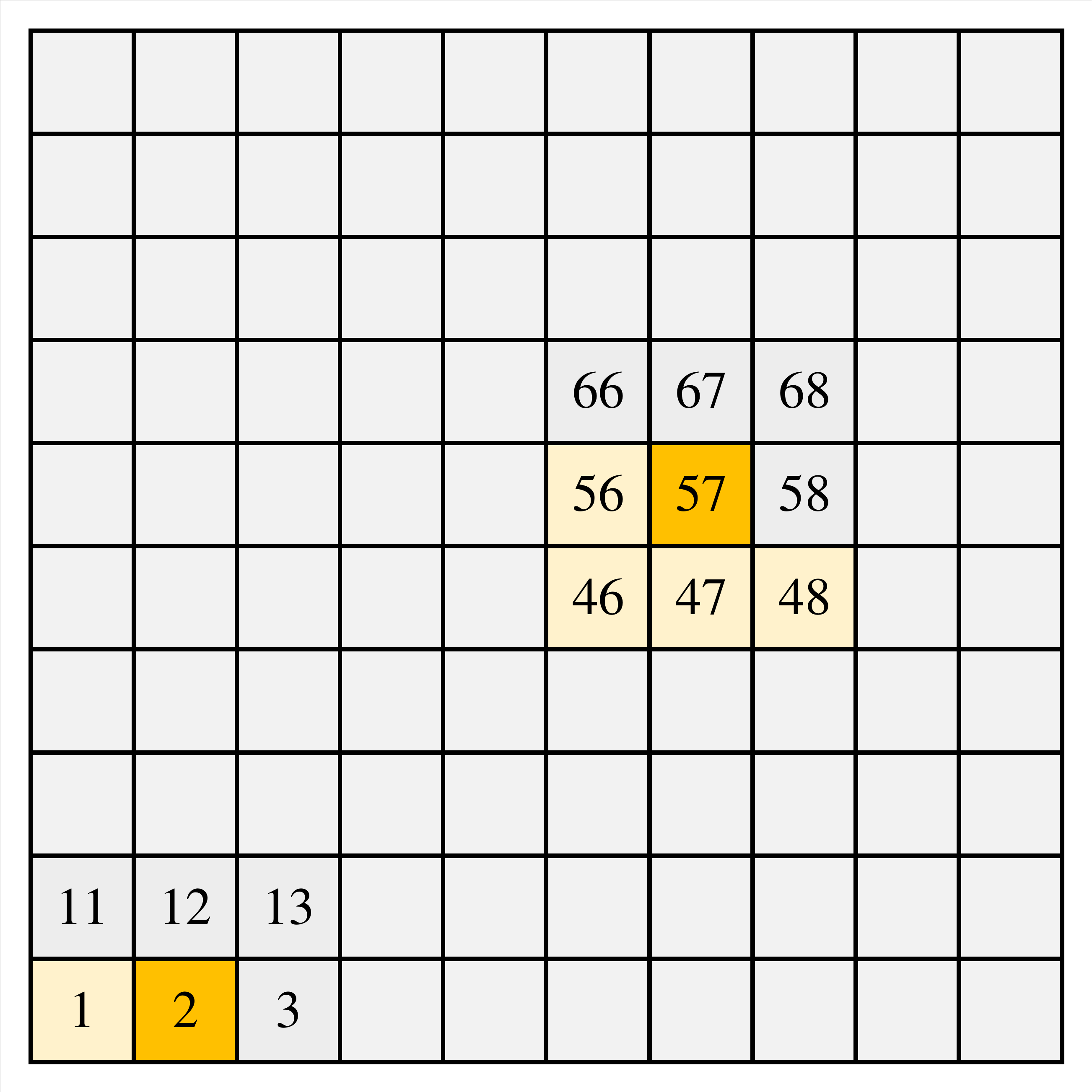}
     }
   \captionsetup{justification=centering}
       \vspace{-0.2cm}
\caption{An example of spatial grid}
\label{fig:spatial-grid}
\end{center}
\end{minipage}
\label{fig:k}
\end{figure*}

Fig.~\ref{fig:spatial-grid}(a) shows how to generate candidate pairs based on spatial grid. The number in a cell is the cell id. We assume that a geo-image is located in the cell 57 colored by yellow, denoted $C_{57}$. In order to retrieve the candidate pairs $(I,\hat{I})$, just only the $C_{57}$ and its eight neighbor cells colored by light yellow need to be accessed due to the restriction of geographical similarity threshold. Therefore, for one geo-image, we only need to check total nine cells to find its partner to form a candidate pair. If the current accessed cell is near the edge of the grid, such as $C_2$, only six cells should be checked for candidates searching. Thus, using this strategy can reduce the size of search space significantly. We utilize the spatial similarity filter to find the result from these cells mentioned above.

\noindent\textbf{SVS-JOIN$_G$ Algorithm}. Based on spatial grid method, we develop a extension of SVS-JOIN$_B$ called SVS-JOIN$_G$ algorithm. Like SVS-JOIN$_B$, this algorithm utilize spatial grid strategy. In other words, a spatial grid is constructed for the input dataset $\mathcal{R}$ as the basic spatial data structure. The geo-images in $\mathcal{R}$ are then accessed in the ascending order of their cell id. For each cell $C_i$, this algorithm will get a cells set denoted as $M[C_i]$, in which the geo-image will be joined with all of the geo-images in $C_i$. In $M[C_i]$, the neighbor cells of $C_i$ have smaller id than $C_i$ itself. 

There are some differences between SVS-JOIN$_B$ and SVS-JOIN$_G$. For example, SVS-JOIN$_G$ algorithm constructs an inverted index for all cells in the grid, rather than a global index. Therefore, for each visual word $v$ in the global visual dictionary, every cell has its inverted index $C_i.\mathcal{L}_v$.

\begin{algorithm}
\begin{algorithmic}[1]
\footnotesize
\caption{\bf SVS-JOIN$_G$ Algorithm}
\label{alg:svs-joins_g}

\INPUT $\mathcal{R}$ is a geo-image dataset sorted by a global ordering $\Phi$, $\Gamma_V$ is the visual similarity threshold, $\Gamma_G$ is a geographical similarity threshold.
\OUTPUT $\mathcal{P}$ is the result pairs set.

\STATE $G(\mathcal{R}) \leftarrow GridConstructor(\mathcal{R},\Gamma_G)$;
\FOR{each $C_i \in G(\mathcal{R})$}
    \STATE $M[C_i] \leftarrow GetJoinCells(G(\mathcal{R}),C_i)$;
    \FOR{each $C_j \in G(\mathcal{R})$}
        \STATE $\mathcal{P} \leftarrow \mathcal{P} \cup \mbox{SVS-JOIN}_B(C_i,C_j,\Gamma_G,\Gamma_V)$;
    \ENDFOR
\ENDFOR
\RETURN $\mathcal{P}$;
\end{algorithmic}
\end{algorithm}

Algorithm~\ref{alg:svs-joins_g} demonstrates the computation process of SVS-JOIN$_G$ algorithm. Similar to SVS-JOIN$_B$ algorithm, the input consists of a geo-image dataset $\mathcal{R}$ sorted by $\Phi$, a visual similarity threshold $\Gamma_V$ and a geographical similarity threshold $\Gamma_G$. The first step is to construct a spatial grid $G(\mathcal{R})$ for $\mathcal{R}$, shown as Line 1. The geo-images are ordered according to cell id and $|I.V|$. After this step, it traverses the $G(\mathcal{R})$ to search the join cell by cell. For each cell $C_i$, the procedure $GetJoinCell(G(\mathcal{R}),C_i)$ is executed to get the cell set $M[C_i]$. For all the cell $C_j \in M[C_j]$, the algorithm executes $\mbox{SVS-JOIN}_B(C_i,C_j,\Gamma_G,\Gamma_V)$ to return the final results set. It is worth noting that the geo-image $I$ located in each cell $C_i$ are checked several times, that means more buffers need to create to store the cells for later processing.

\section{The Quadtree Based Global Index Method}
\label{globalindex}

In the last section, we introduce the baselines which utilize spatial grid to improve the performance of spatial search. For both SVS-JOIN$_B$ and SVS-JOIN$_G$, we build a spatial grid for $\mathcal{R}$ at first, and then construct the local inverted index for each cell of the gird. In this section, we propose a novel method to solve the problem of spatial visual similarity joins efficiently based on a global inverted index and quadtree partition strategy.

\begin{figure*}
\newskip\subfigtoppskip \subfigtopskip = -0.1cm
\begin{minipage}[b]{0.99\linewidth}
\begin{center}
     \subfigure[{A quadtree}]{
     \includegraphics[width=0.57\linewidth]{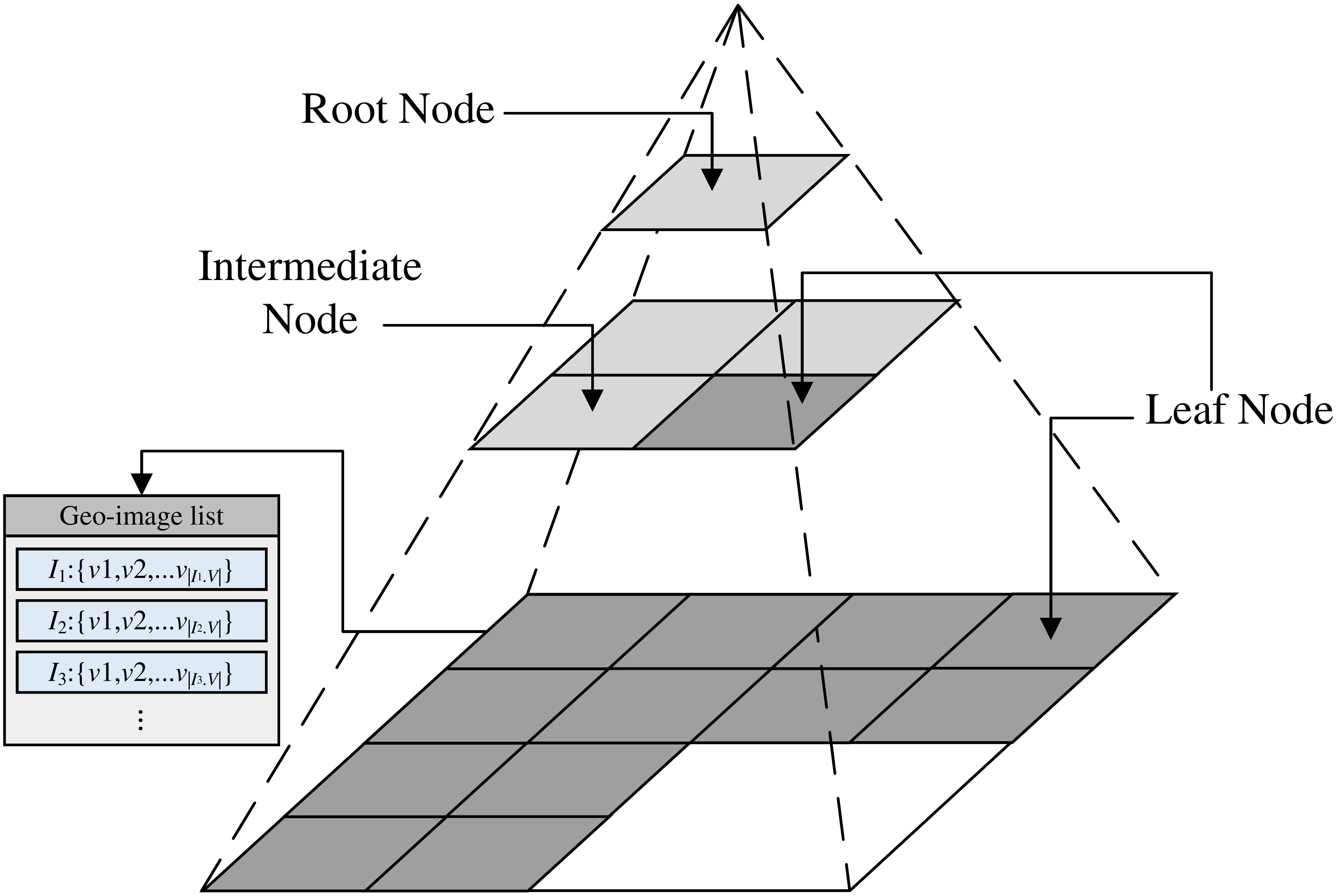}
     }
     \subfigure[{Quadtree partition}]{
     \includegraphics[width=0.39\linewidth]{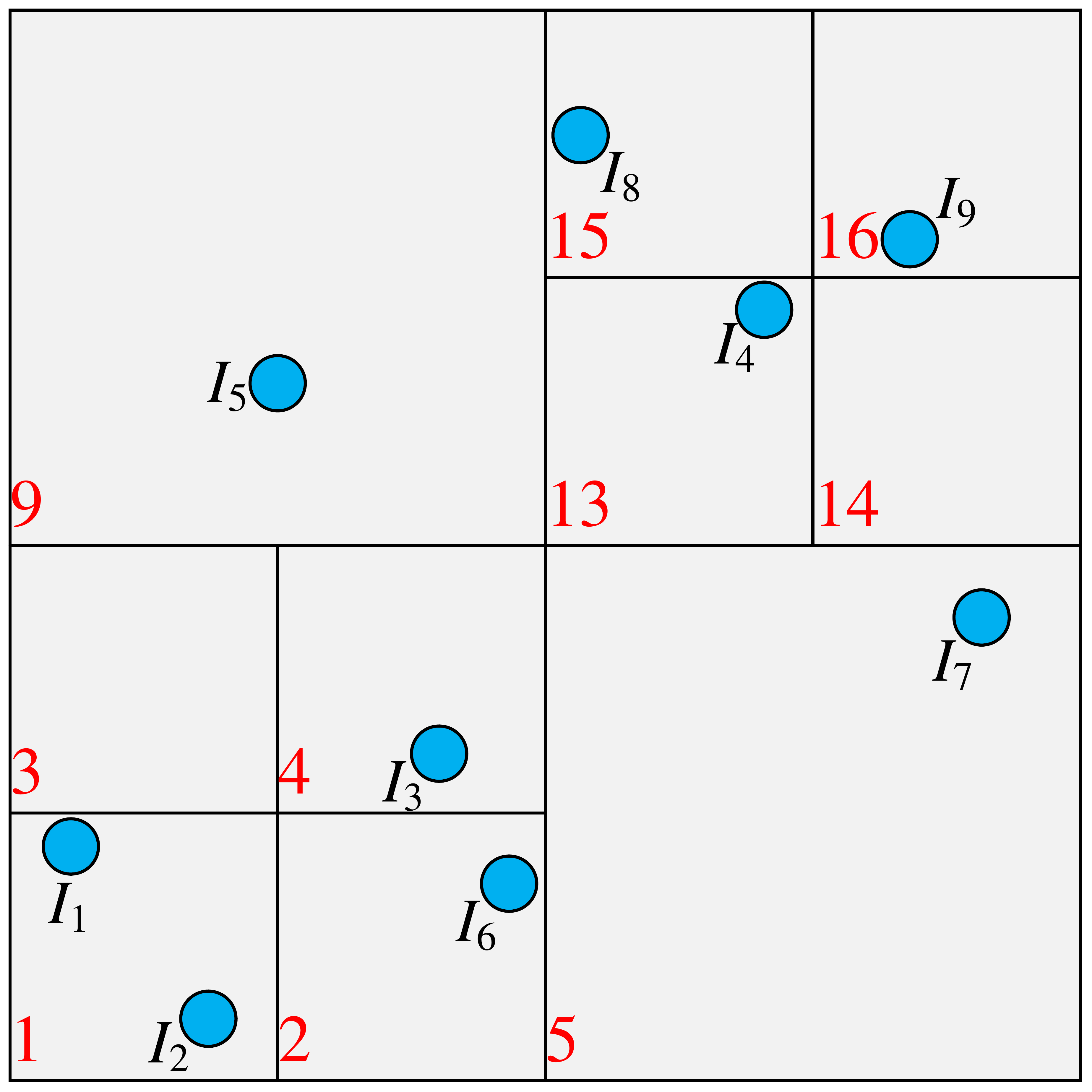}
     }
   \captionsetup{justification=centering}
       \vspace{-0.2cm}
\caption{An example of quadtree partition}
\label{fig:quadtree-partition}
\end{center}
\end{minipage}
\label{fig:k}
\end{figure*}

\subsection{Quadtree Partition and Global Index}
\noindent\textbf{Quadtree Partition}. Quadtree is one of the popular spatial indexing structures used in many applications. It aims to partition a 2-dimensional spatial region into 4 subregions in a recursive manner. Fig.~\ref{fig:quadtree-partition}(a) illustrates an example of quadtree which partitions the spatial region into $L$ levels. For $l$-th level, the region is split into $4^l$ equal subregions. Each node of quadtree corresponds to a subregion. The root node of quadtree locate on the $0$-th level, which represents the whole spatial region. Four subnodes in $1$-level are partitioned from the root node in $0$-th level. And the subnodes in $3$-level are split from the nodes in $2$-level as the same manner. From the Fig.~\ref{fig:quadtree-partition}(a) we can find that there are three colors of nodes. In specific, the light gray nodes are root node and intermediate nodes. The dark gray nodes in any level of the quadtree are the leaf nodes according to the split condition. For each leaf node, there is a list of geo-images in it. In general, the whole spatial region is partitioned into several nodes and the geo-images destribute in these nodes.

Fig.~\ref{fig:quadtree-partition}(b) shows the partition of the Example~\ref{ex:geo-images} by a quadtree. The red color number in quadtree is the node id. Apparently, these 9 geo-images distribute in the subregions. For node 1, denoted as $N_1$, it contains two geo-image $I_1$ and $I_2$. As the number of geo-images in $\mathcal{R}$ is really small, the other nodes contain only one geo-image at most.

\begin{figure*}
\newskip\subfigtoppskip \subfigtopskip = -0.1cm
\begin{minipage}[b]{1\linewidth}
\begin{center}
     \subfigure[{A quadtree}]{
     \includegraphics[width=0.48\linewidth]{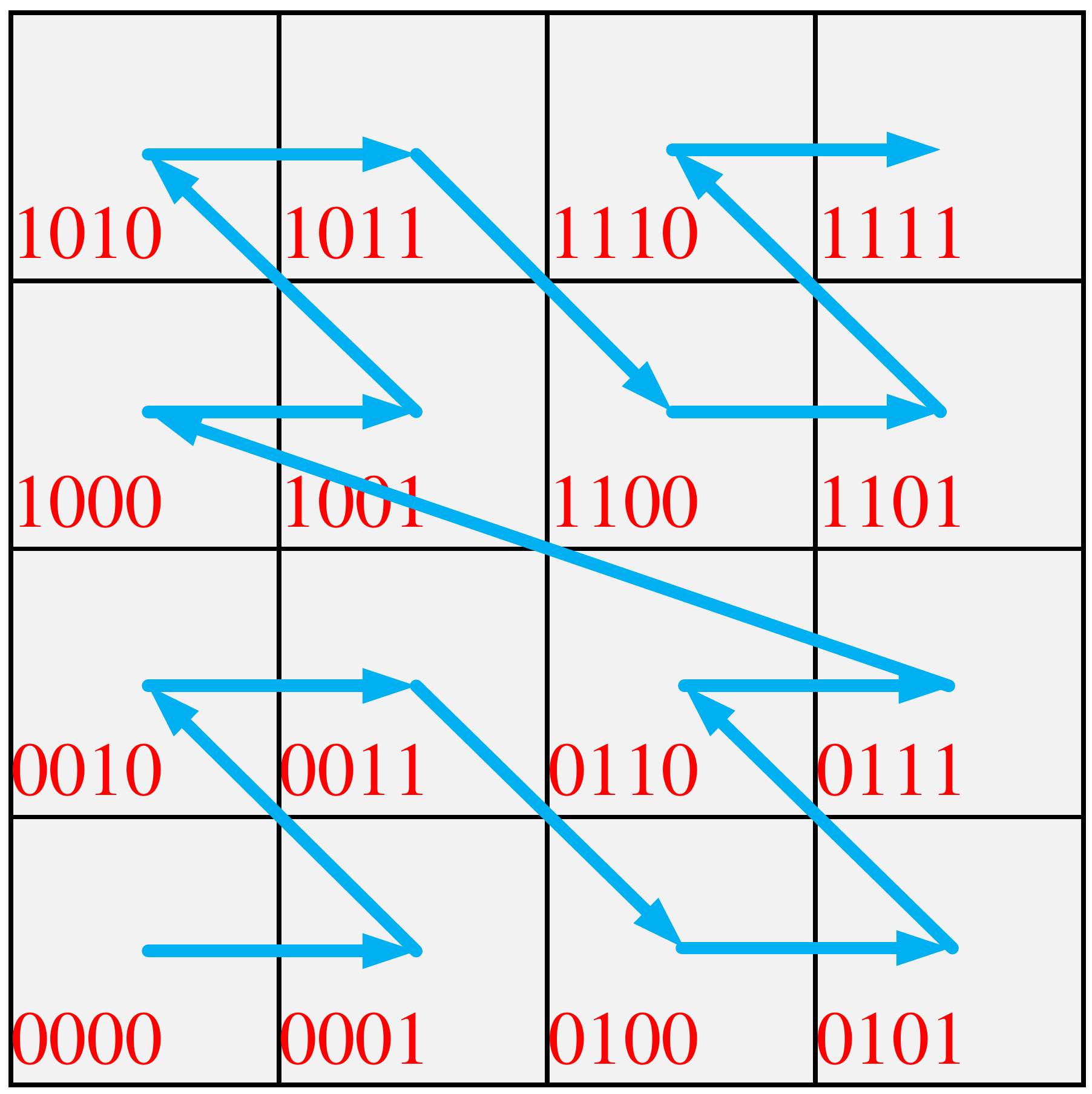}
     }
     \subfigure[{Quadtree partition}]{
     \includegraphics[width=0.48\linewidth]{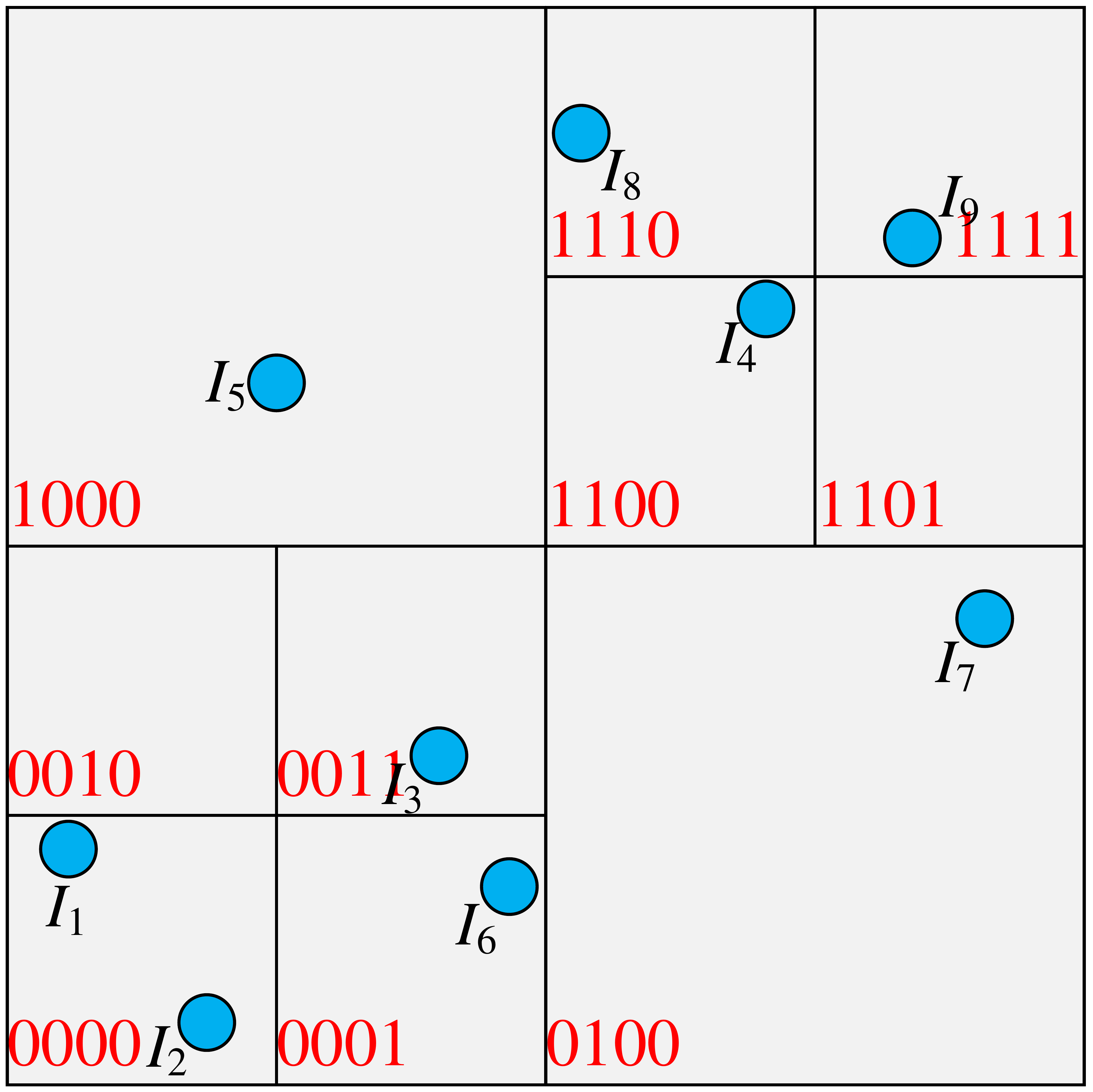}
     }
   \captionsetup{justification=centering}
       \vspace{-0.2cm}
\caption{An example of Z-order}
\label{fig:z-order-curve}
\end{center}
\end{minipage}
\label{fig:k}
\end{figure*}

\noindent\textbf{Z-Order Curve}. In this paper, we utilize Z-order curve to encode the each node of quadtree, which is encoded based on its partition sequence. There is a direct relationship between z-order curve and quadtree. The Z-order curve can describe the path of the node of a quadtree. Fig.~\ref{fig:z-order-curve}(a) demonstrates how to generate the Morton code of a subregion based on spatial partition sequence in a region. According to Z-order curve, we denote these 16 subregions from 0 to 15 in decimal, or from 0000 to 1111 in binary. Fig.~\ref{fig:z-order-curve}(b) illustrates the Morton code in the quadtree partition of Example~\ref{ex:geo-images}, we use the code in binary as the node id.

\subsection{SVS-JOIN$_Q$ Algorithm}
Based on the quadtree partition and the global inverted index, we develop a novel algorithm called SVS-JOIN$_Q$ to solve the spatial visual similarity joins problem efficiently. Algorithm~\ref{alg:svs-joins_q} shows the pseudo-code of this algorithm and Algorithm~\ref{alg:global-index-constructor} and Algorithm~\ref{alg:join-search} demonstrates two key procedures applied in SVS-JOIN$_Q$. The first step of SVS-JOIN$_Q$ is to construct a quadtree to partition the whole spatial region of the input dataset $\mathcal{R}$. After that, it executes the procedure $GlobalIndexConstructor(\hat{\mathcal{R}},\Gamma_V)$ to build the global inverted index for each visual word. When building the inverted index lists for geo-image $I$, only the geo-image $\hat{I}$ in the neighbors nodes or the same node need to be considered. Then for each inverted indexing list, the algorithm recalls the start position and end position of each node and the exact position of geo-images for searching. $JoinSearch(I,\mathcal{I}_G,\Gamma_G,\Gamma_V)$ is invoked to search out all the similar geo-image pairs as the result. 

\begin{algorithm}
\begin{algorithmic}[1]
\footnotesize
\caption{\bf SVS-JOIN$_Q$ Algorithm}
\label{alg:svs-joins_q}

\INPUT a geo-image dataset $\mathcal{R}$, a visual similarity threshold $\Gamma_V$, a geographical similarity threshold $\Gamma_G$.
\OUTPUT a result pairs set $\mathcal{P}$.

\STATE $T_{quad} \leftarrow QuadtreeConstructor(\mathcal{R},\Gamma_G)$;
\STATE $\hat{\mathcal{R}} \leftarrow AscSortZ(\mathcal{R})$;
\STATE $\mathcal{I}_G \leftarrow GlobalIndexConstructor(\hat{\mathcal{R}},\Gamma_V)$;
\FOR{each $I \in \hat{\mathcal{R}}$}
    \STATE $\mathcal{P} \leftarrow \mathcal{R} \cup JoinSearch(I,\mathcal{I}_G,\Gamma_V,\Gamma_G)$;
\ENDFOR
\RETURN $\mathcal{P}$;
\end{algorithmic}
\end{algorithm}

\begin{algorithm}
\begin{algorithmic}[1]
\footnotesize
\caption{\bf GlobalIndexConstructor($\hat{\mathcal{R}}$,$\Gamma_V$)}
\label{alg:global-index-constructor}

\INPUT a geo-image dataset $\hat{\mathcal{R}}$, a visual similarity threshold $\Gamma_V$.
\OUTPUT a inverted index set $\mathcal{I}_G$.

\STATE Initializing: sort the visual words in descending order of number of non-zero entries;
\STATE Initializing: Denote the maximum of $w(I.V[i])$ for all $I \in \mathcal{R}$ as maxweight of $i$-th visual word;
\STATE Initializing: Denote the maximum of $w(I.V[i])$ from 1 to $m$ as maxweight of $I.V$;
\STATE Initializing: $\mathcal{P} \leftarrow \emptyset$;
\STATE Initializing: $\forall \iota_i \in \mathcal{I}_G \leftarrow \emptyset$;
\STATE Initializing: $S_V \leftarrow \emptyset$;

\FOR{each $I \in \hat{\mathcal{R}}$}
    \STATE $\beta \leftarrow$  set of geo-iamge in $I.node$ or $I.neighbors$;
    \STATE Denote the maximum of $I.V[i]$ for all $I \in r$ as maxweight$_i(r)$;
    \FOR{each $I.V[i] > 0$ in ascending order of $i$}
        \STATE $S_V \leftarrow S_V +$ maxweight$_i(r)*I.V[i]$;
        \IF{$S_V > \Gamma_V$}
            \STATE $InvertedIndexConstructor(\iota_i)$;
        \ENDIF
    \ENDFOR
\ENDFOR
\FOR{each $\iota_j \in \mathcal{I}_G$}
    \STATE Record $p_{start}$ and $p_{end}$ of each node in $\iota_j$;
    \STATE Record the $p_{I_i}$ in $\iota_j$
\ENDFOR
\RETURN $\mathcal{I}_G$;
\end{algorithmic}
\end{algorithm}

\begin{algorithm}
\begin{algorithmic}[1]
\footnotesize
\caption{\bf JoinSearch($I$,$\mathcal{I}_G$,$\Gamma_G$,$\Gamma_V$)}
\label{alg:join-search}

\INPUT a geo-image $I$, a global index set $\mathcal{I}_G$, a visual similarity threshold $\Gamma_V$, a geographical similarity threshold $\Gamma_G$.
\OUTPUT a result pairs set $\mathcal{P}$.

\STATE Initializing: $S_T \leftarrow \emptyset$;
\STATE Initializing: $\mathcal{P} \leftarrow \emptyset$;
\STATE Initializing: $score \leftarrow \sum_{i}^{}I.V[i]*$maxweight$_i(\mathcal{R})$;

\FOR{each $i$ s.t. $I.V[i] > 0$}
    \FOR{each node $\mathcal{N} \in I.N \cup I.neighbors$}
        \STATE $S_V \leftarrow S_V +$ maxweight$_i(r)*I.V[i]$;
        \IF{$\mathcal{N} \in I.neighbors$}
            \STATE $p_{start} = \mathcal{N}.p_{start}$;
            \STATE $p_{end} = \mathcal{N}.p_{end}$;
        \ELSE
            \STATE $p_{start} = GetPosition(\iota_i, I)$;
            \STATE $p_{end} = \mathcal{N}.p_{end}$;
        \ENDIF
        \FOR{each $\hat{I} \in \iota_i[p_{start},p_{end}]$}
            \IF{$I$ equals $\hat{I}$}
                \STATE Continue;
            \ENDIF
            \IF{$Sim[\hat{I}] \neq 0 \ || \ score \leq \Gamma_G$}
                \STATE $Sim[\hat{I}] \leftarrow Sim[\hat{I}]+I.V[i]*\hat{I}.V[i]$;
            \ENDIF
            \STATE $score \leftarrow score-I.V[i]*$maxweight$_i(\mathcal{R})$;
        \ENDFOR
    \ENDFOR
\ENDFOR
\STATE $Verify(I,\hat{I},S_V,\mathcal{P})$;
\RETURN $\mathcal{P}$;
\end{algorithmic}
\end{algorithm}

\section{PERFORMANCE EVALUATION}
\label{perform}

In this section, we present results of a comprehensive performance evaluation on real geo-image datasets to evaluate the efficiency and scalability of the proposed approaches. Specifically, we evaluate the efficiency of the following methods.

\begin{itemize}
\item \textbf{SVS-JOIN$_B$}. SVS-JOIN$_B$ is the technique introduced in Section~\ref{base}.
\item \textbf{SVS-JOIN$_G$}. SVS-JOIN$_G$ is the technique introduced in Section~\ref{base}.
\item \textbf{SVS-JOIN$_Q$}. SVS-JOIN$_Q$ is the technique introduced in Section~\ref{globalindex}.
\end{itemize}

\noindent\textbf{Datasets.} Performance of three algorithms is evaluated on both real spatial and image datasets.
The following two datasets are deployed in our experiments. Real image dataset \textbf{Flickr} is obtained by crawling millions image from the popular photo-sharing platform Flickr(\url{http://www.flickr.com/}). To evaluate the scalability of our proposed algorithm, The dataset size varies from 100K to 500K. The geo-location information can be obtained from the geo-tag of each image. Similarly, Real dataset \textbf{ImageNet} is obtained from is the largest image dataset ImageNet, which is widely used in image processing and computer vision. it includes 14,197,122 images and 1.2 million images with SIFT features. We generate \textbf{ImageNet} dataset with varying size from 100K to 500K. The geographical information of the images are randomly generated from spatial datasets Rtree-Portal (\url{http://www.rtreeportal.org}).

\noindent\textbf{Workload}. The geo-image dataset size increases from 100K to 500K; the number of the visual words contained in a geo-image grows from 20 to 100; the geographical similarity threshold $\Gamma_G$ and visual similarity threshold $\Gamma_V$ varies from 0.02 to 0.10 and from 0.5 to 0.9 respectively. By default, The image dataset size, the number of the visual words, the geographical similarity threshold, visual similarity threshold set to \textbf{300K}, \textbf{60}, \textbf{0.06}, \textbf{0.7} respectively.

All the Experiments are run on a PC with Intel(R) Xeon 2.60GHz dual CPU and 16G memory running Ubuntu 16.04 LTS. All algorithms in the experiments are implemented in Java. Note that the quadtree of SVS-JOIN$_Q$ method is maintained in memory.

\begin{figure*}
\newskip\subfigtoppskip \subfigtopskip = -0.1cm
\begin{minipage}[b]{1\linewidth}
\begin{center}
     \subfigure[Evaluation on Flickr]{
     \includegraphics[width=0.48\linewidth]{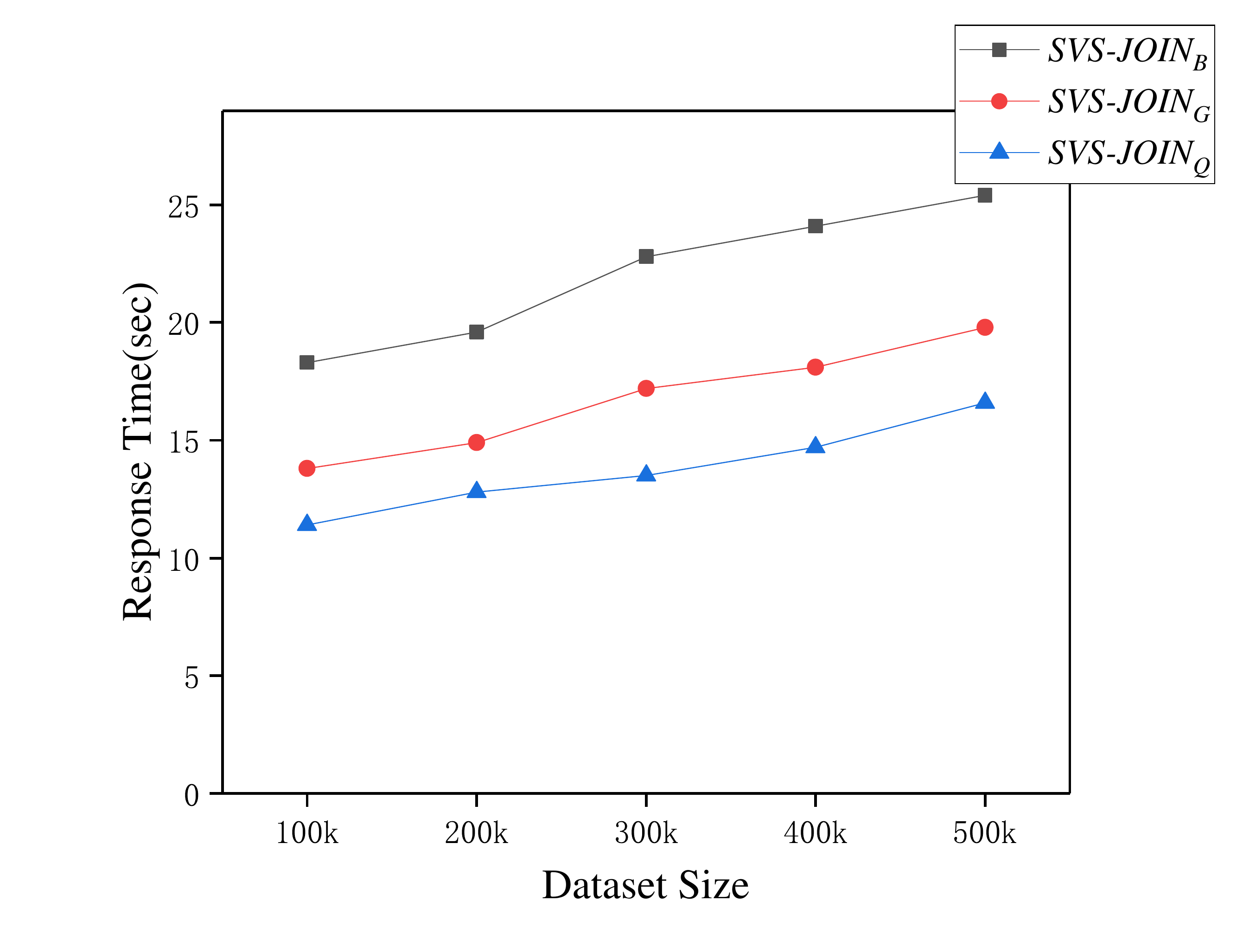}
     }
     \subfigure[Evaluation on ImageNet]{
     \includegraphics[width=0.48\linewidth]{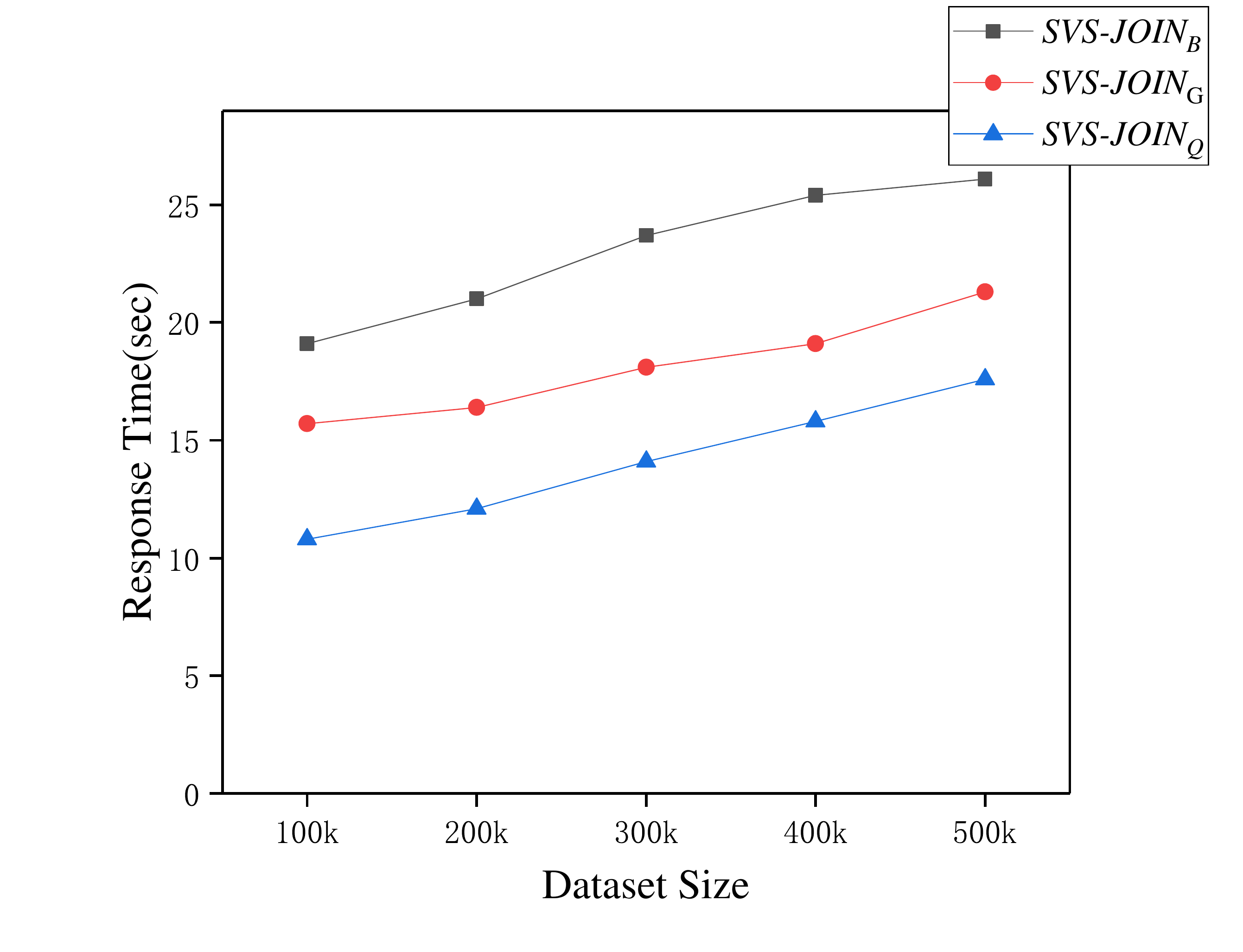}
     }
   \captionsetup{justification=centering}
       \vspace{-0.2cm}
\caption{Evaluation on various dataset size on Flickr and ImageNet}
\label{fig:dataset-size}
\end{center}
\end{minipage}
\end{figure*}

\noindent\textbf{Evaluation on the size of dataset.} We evaluate the effect of the size of dataset on Flickr and ImageNet shown in Fig.~\ref{fig:dataset-size}. It is obvious that the response time of SVS-JOIN$_B$, SVS-JOIN$_G$ and SVS-JOIN$_Q$ increase gradually in Fig.~\ref{fig:dataset-size}(a). Specifically, The performance of SVS-JOIN$_B$ is the worse than two others as no effective spatial search technique in this solution. The time cost of SVS-JOIN$_G$ fluctuate from about 14 second to 23 second, which is higher than SVS-JOIN$_Q$ because the quadtree and global inverted index based solution is more efficient. Fig.~\ref{fig:dataset-size}(b) illustrates that the evaluation on ImageNet dataset. Similar to the situation on Flickr dataset, the efficiency of SVS-JOIN$_Q$ is the highest. However, with the rising of the dataset size from 100k to 500k, the speed of increment of time cost of SVS-JOIN$_Q$ is higher than the speed on Flickr. On the other hand, the performance of SVS-JOIN$_B$ is still the worst.

\begin{figure*}
\newskip\subfigtoppskip \subfigtopskip = -0.1cm
\begin{minipage}[b]{1\linewidth}
\begin{center}
     \subfigure[Evaluation on Flickr]{
     \includegraphics[width=0.48\linewidth]{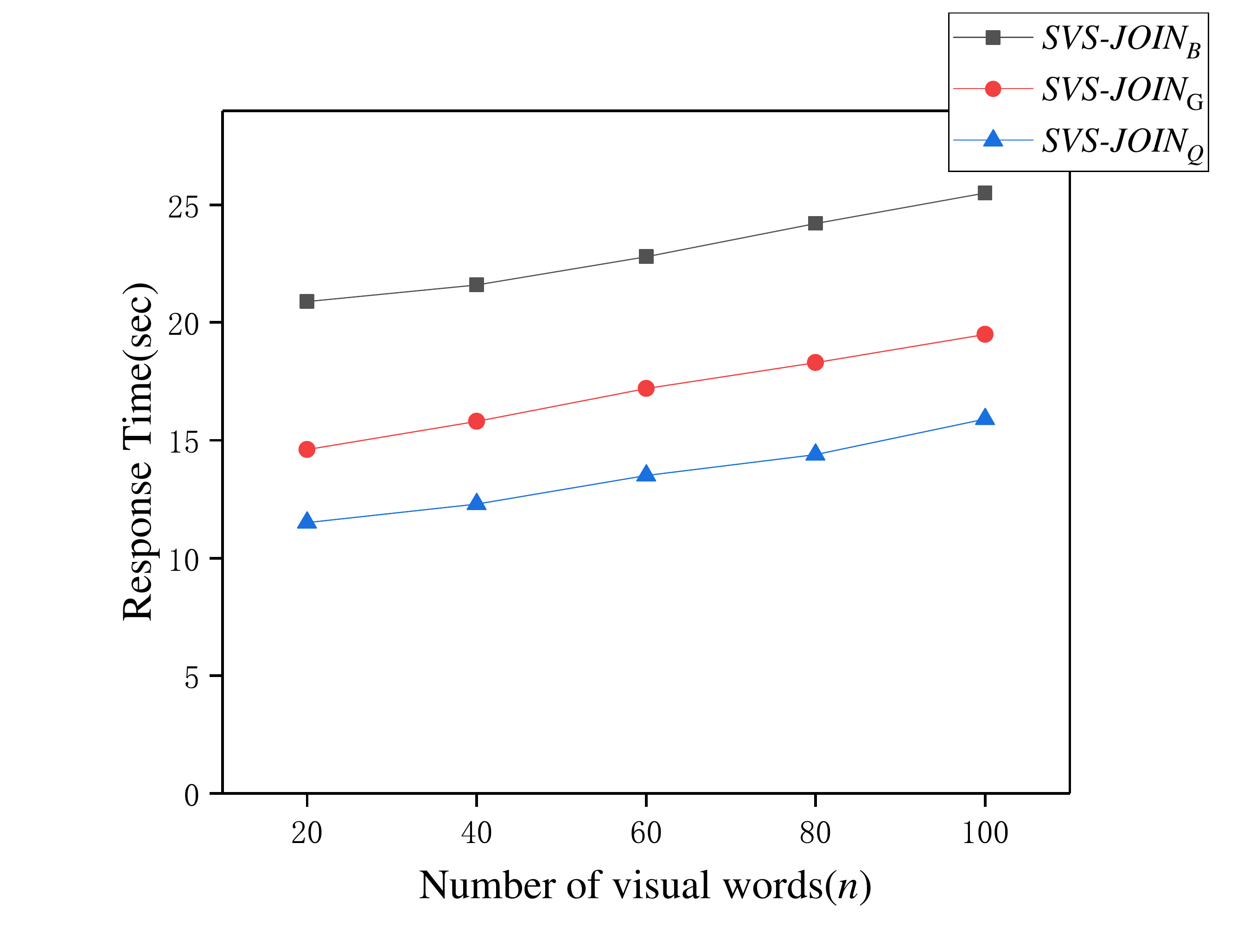}
     }
     \subfigure[Evaluation on ImageNet]{
     \includegraphics[width=0.48\linewidth]{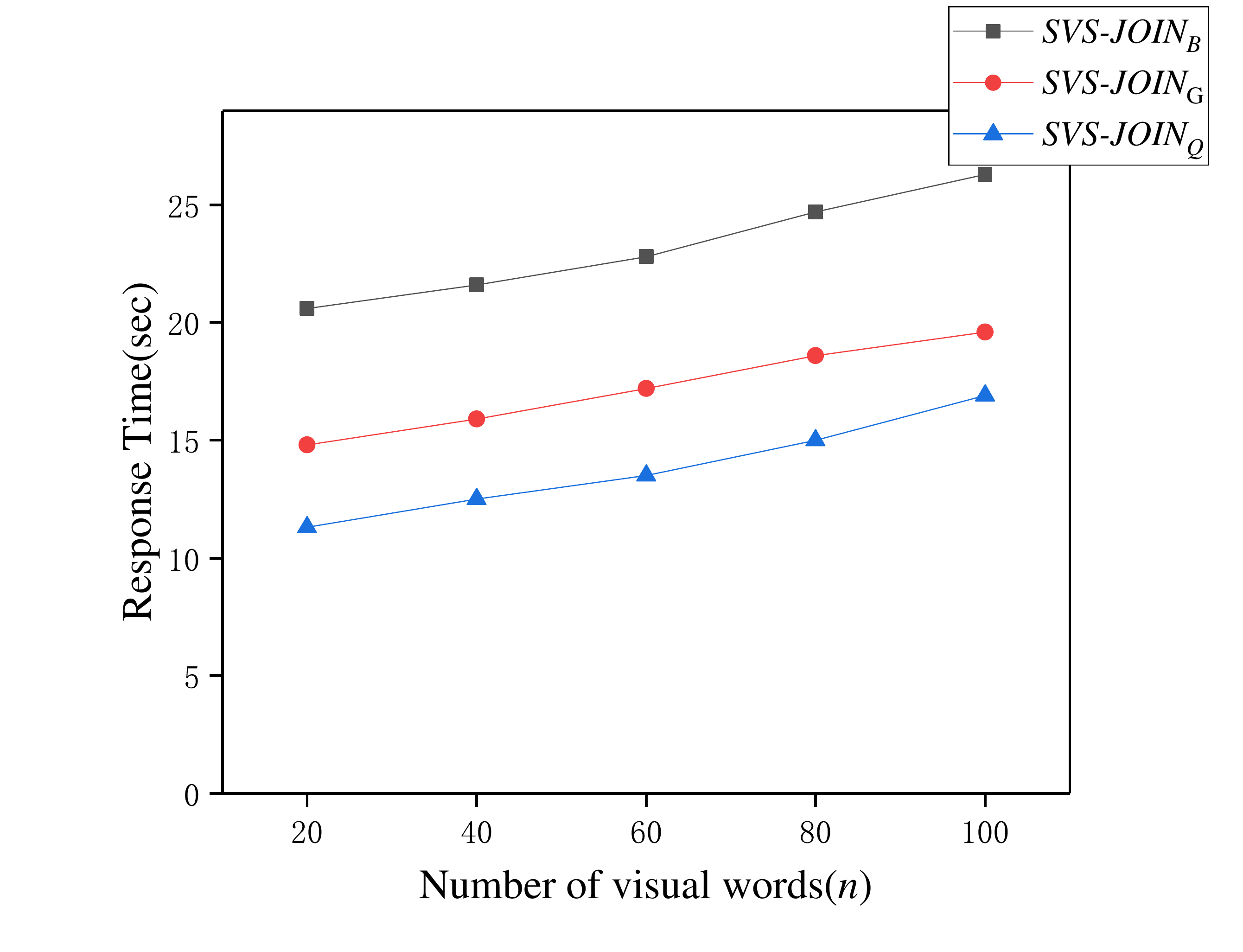}
     }
   \captionsetup{justification=centering}
       \vspace{-0.2cm}
\caption{Evaluation on the number of visual words on Flickr and ImageNet}
\label{fig:number-of-query-visual-words}
\end{center}
\end{minipage}
\end{figure*}

\noindent\textbf{Evaluation on the number of visual words.} We evaluate the effect of the number of visual words on Flickr and ImageNet dataset shown in Figure~\ref{fig:number-of-query-visual-words}. We can see from Fig.~\ref{fig:number-of-query-visual-words}(a) that the response time of all these three methods grow step by step with the increment of number of visual words. For SVS-JOIN$_B$, when the number of visual words is larger than 40, the growth speed of it is a little faster. Apparently, the response time of it is the highest of them. SVS-JOIN$_Q$ is the most efficient algorithm among them on this dataset. The evaluation on ImageNet dataset is shown in Fig.~\ref{fig:number-of-query-visual-words}(b). In the interval $[60,100]$ the growth speed of SVS-JOIN$_B$ and SVS-JOIN$_Q$ are faster. However, this situation does not appear in SVS-JOIN$_G$. There is no doubt the performance of SVS-JOIN$_Q$ is the best, just like the evaluations mentioned above.

\begin{figure*}
\newskip\subfigtoppskip \subfigtopskip = -0.1cm
\begin{minipage}[b]{1\linewidth}
\begin{center}
     \subfigure[Evaluation on Flickr]{
     \includegraphics[width=0.48\linewidth]{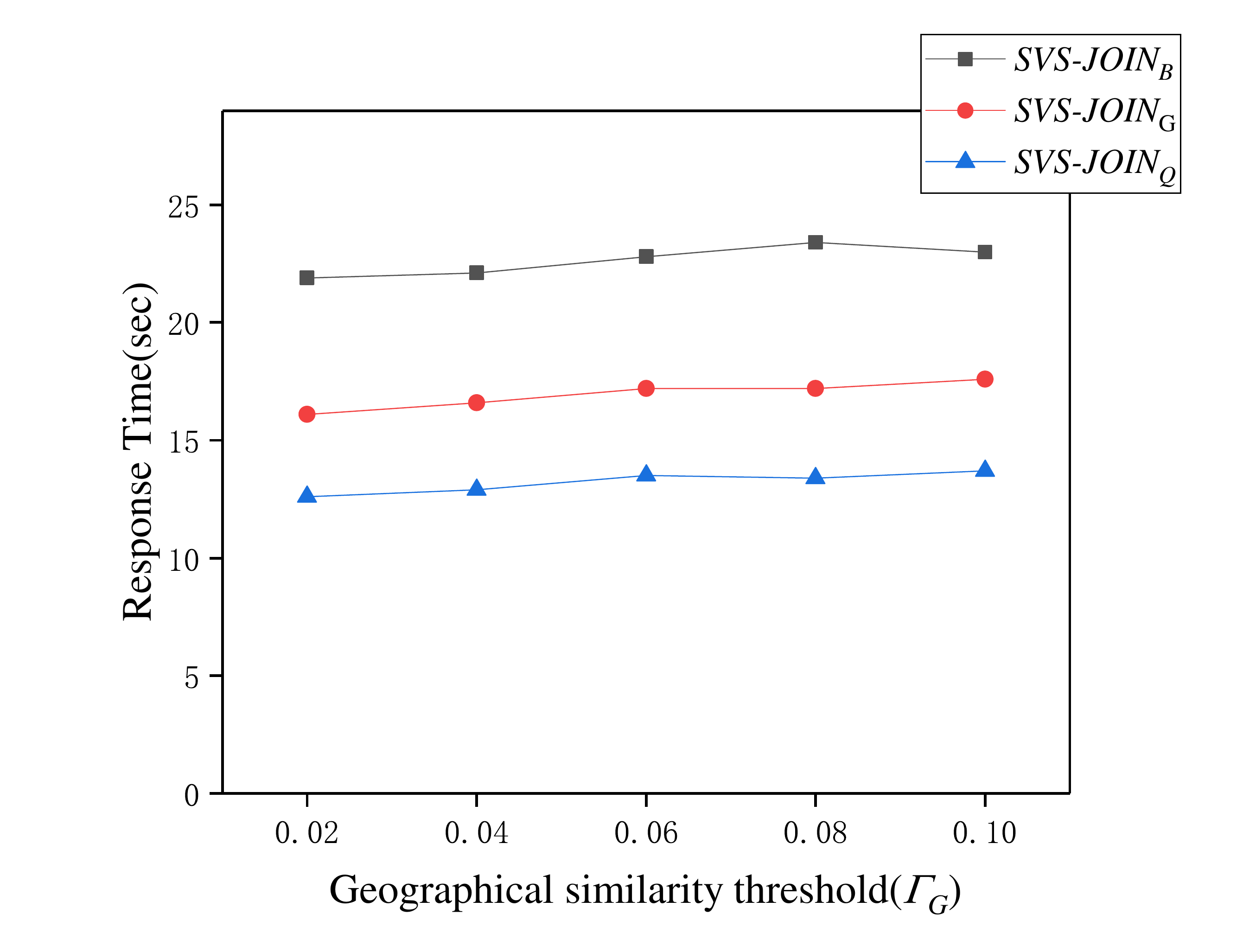}
     }
     \subfigure[Evaluation on ImageNet]{
     \includegraphics[width=0.48\linewidth]{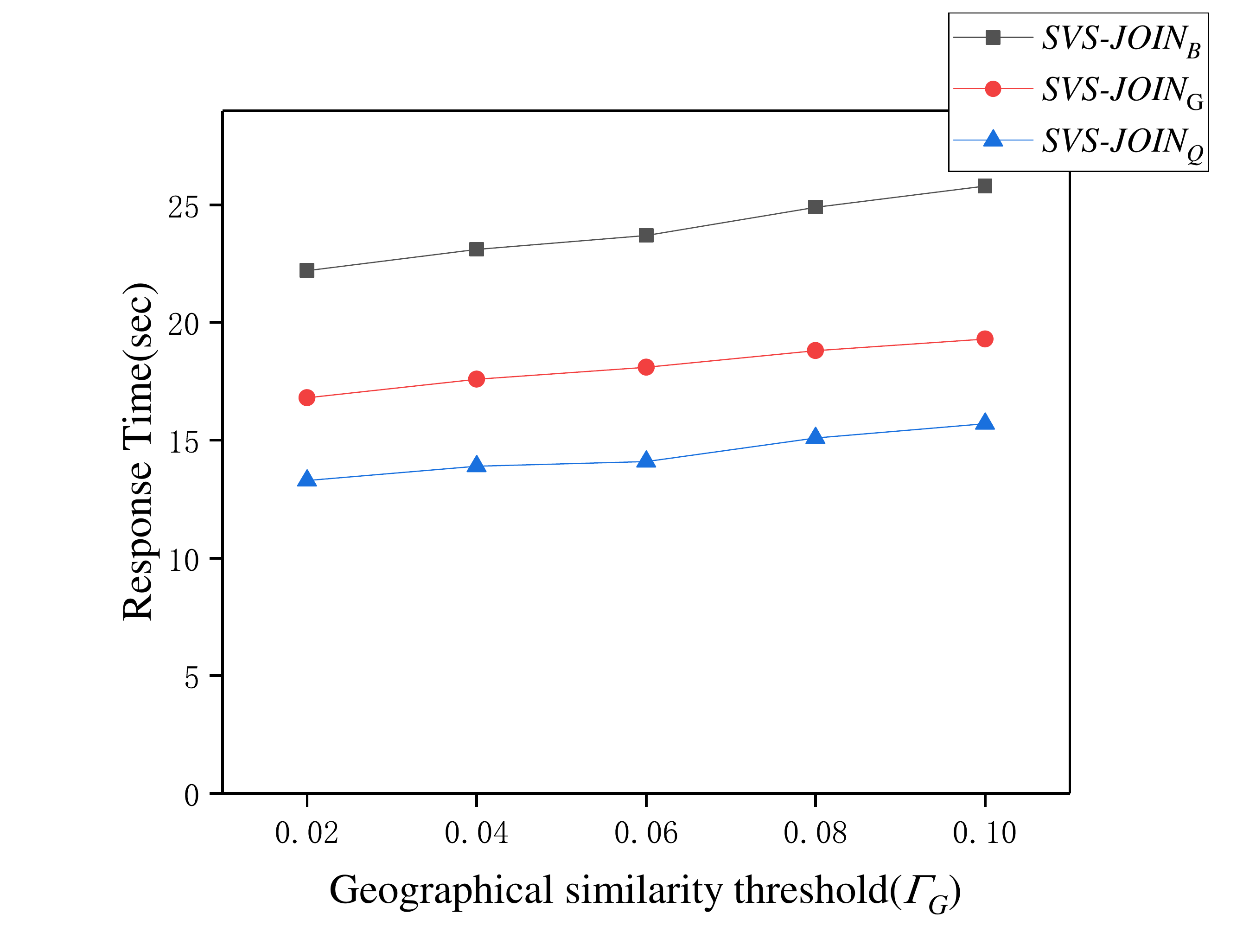}
     }
   \captionsetup{justification=centering}
       \vspace{-0.2cm}
\caption{Evaluation on the geographical similarity threshold on Flickr and ImageNet}
\label{fig:geo-response-time}
\end{center}
\end{minipage}
\end{figure*}

\noindent\textbf{Evaluation on the geographical similarity threshold.} We evaluate the effect of the spatial similarity threshold on Flickr and ImageNet dataset shown in Figure~\ref{fig:geo-response-time}. In Figure~\ref{fig:geo-response-time}(a), with the increasing of geographical similarity threshold, the response time of all these three algorithms are almost unchanged. For SVS-JOIN$_B$, the time cost of it is slightly fluctuated in the interval $[0.02,0.10]$, which is the highest among them. For SVS-JOIN$_Q$ algorithm, the range of its fluctuation is very small, and this method has the lowest response time. On the other hand, the trend of SVS-JOIN$_G$ is similar to SVS-JOIN$_Q$, although its efficiency is not higher than the other one. We can find from Fig.~\ref{fig:geo-response-time}(b) that on ImageNet dataset, The trend of them is slightly different from the situation on Flickr. In specific, when threshold $\Gamma_G \geq 0.06$, both of SVS-JOIN$_B$ and SVS-JOIN$_Q$ have a obvious rise. However, SVS-JOIN$_G$ algorithm seems to be unaffected by the increasing of $\Gamma_G$.

\begin{figure*}
\newskip\subfigtoppskip \subfigtopskip = -0.1cm
\begin{minipage}[b]{1\linewidth}
\begin{center}
     \subfigure[Evaluation on Flickr]{
     \includegraphics[width=0.48\linewidth]{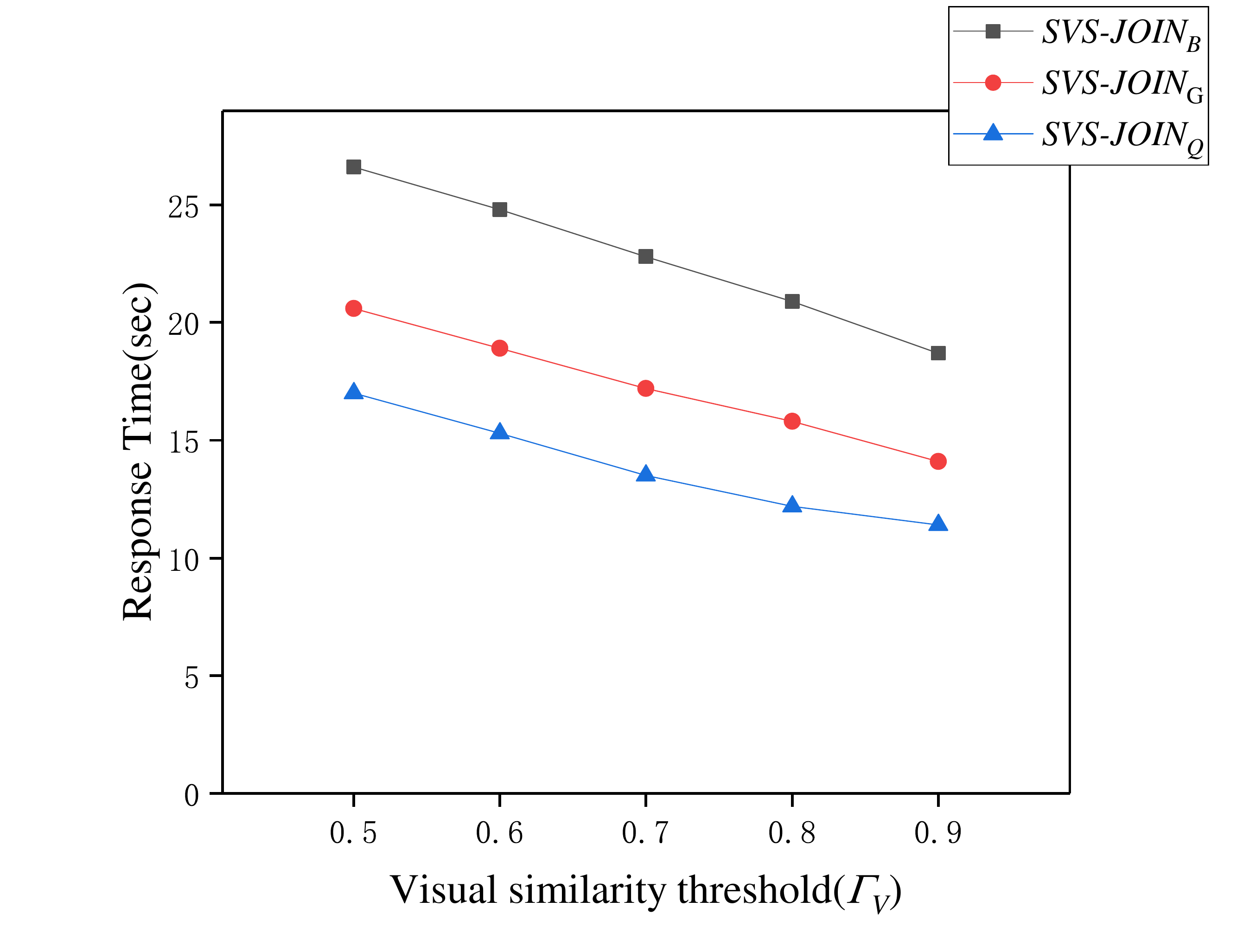}
     }
     \subfigure[Evaluation on ImageNet]{
     \includegraphics[width=0.48\linewidth]{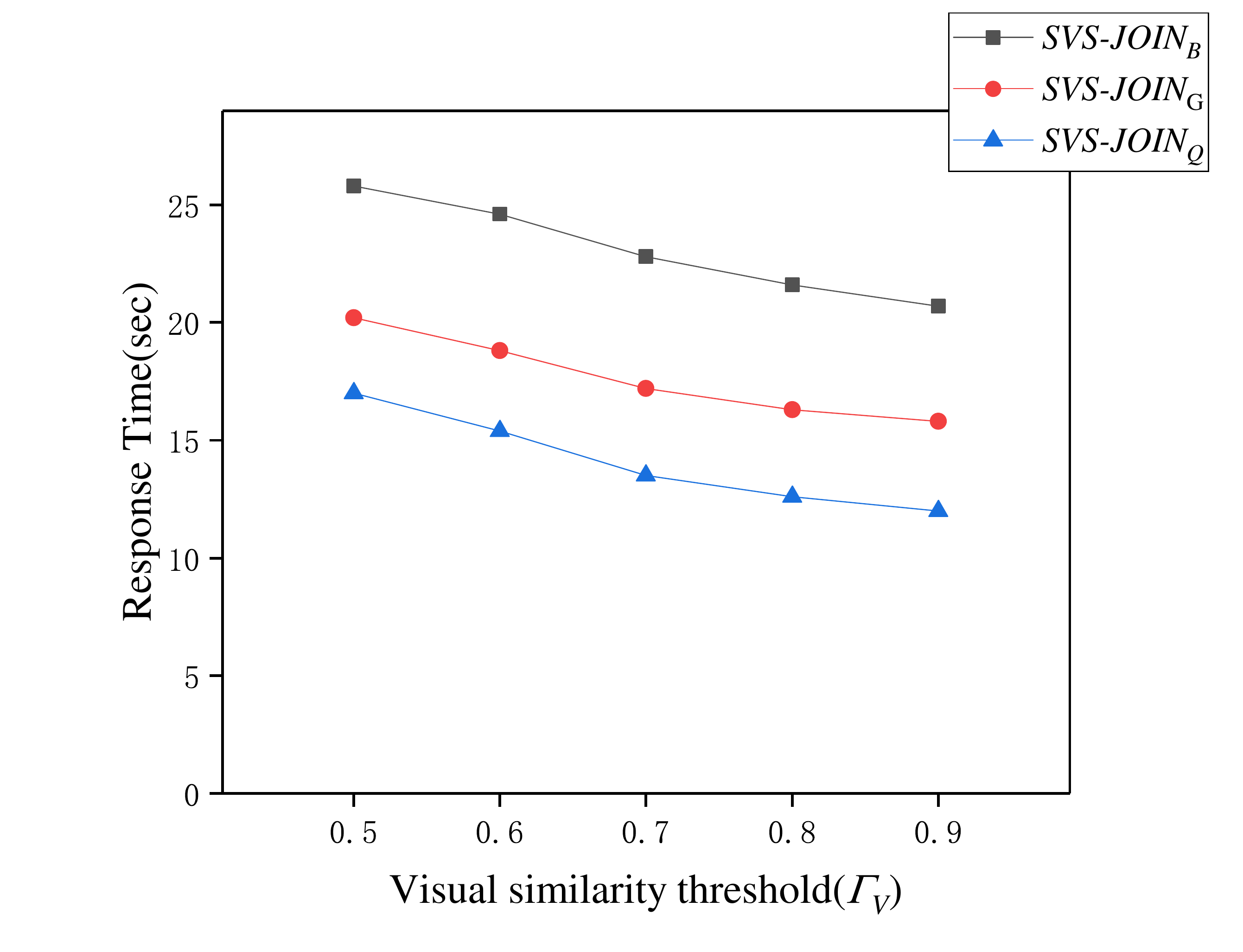}
     }
   \captionsetup{justification=centering}
       \vspace{-0.2cm}
\caption{Evaluation on the visual similarity threshold on Flickr and ImageNet}
\label{fig:visual-response-time}
\end{center}
\end{minipage}
\end{figure*}

\noindent\textbf{Evaluation on the visual similarity threshold.} We evaluate the effect of the visual similarity threshold on Flickr and ImageNet dataset shown in Figure~\ref{fig:visual-response-time}. We can see from the Figure~\ref{fig:visual-response-time}(a) that with the rising of visual similarity threshold, the performance of these three algorithm increase rapidly. The efficiency of SVS-JOIN$_Q$ is higher than SVS-JOIN$_B$ and SVS-JOIN$_G$ from start to finish. In Fig.~\ref{fig:visual-response-time}(b). The response time of them decline gradually but the speed of decrement is a litter slow than the speed on Flickr dataset. Same as the situation above, the performance of SVS-JOIN$_Q$ is the highest.

\section{Conclusion}
\label{con}

In this paper, we study a novel problem named spatial visual similarity joins (SVS-JOIN for short). Given a set of geo-images which contains geographical information and visual content information, SVS-JOIN aims to search out all the geo-image pairs from the dataset, which are similar to each other in both aspects of geographical similarity and visual similarity. To solve this problem efficiently, we define SVS-JOIN in formal at first time and then propose the geographical and visual similarity function. A baseline named SVS-JOIN$_B$ is developed by us inspired from the approaches applied on spatial similarity joins. In order to improve the efficiency of searching, we extent this method and propose a novel algorithm called SVS-JOIN$_G$ which utilizes spatial grid strategy to enhance the performance of spatial retrieval. Besides, we introduce an alternative algorithm named SVS-JOIN$_B$ which applies quadtree tchnique and a global inverted indexing structure. The experimental evaluation on real geo-multimedia dataset shows that our method has a really high performance.

\textbf{Acknowledgments:} This work was supported in part by the National Natural Science Foundation of China
(61702560), project (2018JJ3691, 2016JC2011) of Science and Technology Plan of Hunan Province, and the Research and Innovation Project of Central South University Graduate Students(2018zzts177,2018zzts588).

\bibliographystyle{spmpsci}      
\bibliography{ref}

\end{document}